\documentclass[sigconf]{acmart}
\usepackage{wrapfig,stfloats}
\usepackage{subcaption}
\setcopyright{none}
\renewcommand\footnotetextcopyrightpermission[1]{}
\usepackage[utf8]{inputenc} %
\usepackage[T1]{fontenc}    %
\usepackage{hyperref}       %
\usepackage{url}            %
\usepackage{booktabs}       %
\usepackage{amsfonts}       %
\usepackage{nicefrac}       %
\usepackage{microtype}      %
\usepackage{xcolor}         %
\usepackage{graphicx}
\usepackage{xspace}
\usepackage{multirow}
\usepackage{tikz}
\usepackage{comment}
\usepackage{enumitem}
\usepackage{makecell}       % 添加makecell包用于表格中的单元格格式控制
\usepackage{algorithm}
\usepackage{algpseudocode}
\usepackage{listings}
\usepackage[most]{tcolorbox}  % 添加tcolorbox包用于创建彩色文本框

\newtcolorbox[auto counter]{response}[1][]{
    enhanced,
    top=15pt,
    bottom=15pt,
    left=25pt,
    right=25pt,
    left skip=30pt,
    right skip=30pt,
    colback=gray!5,
    colframe=black,
    fonttitle=\bfseries,
    coltitle=white,
    title=Prompt~\thetcbcounter: #1
}

\usepackage{url}
\usepackage{graphicx}
\usepackage{xspace}
\usepackage{colortbl}
\usepackage{booktabs}
\usepackage{multirow,marvosym}

\usepackage{xcolor}

\definecolor{NegativeRed}{RGB}{220,20,60}
\definecolor{ForestGreen}{rgb}{0, 0.69, 0.31}
\definecolor{NavyBlue}{rgb}{0, 0.44, 0.75}
\usepackage{pifont}

\definecolor{Cerulean}{rgb}{0.0, 0.48, 0.65}

\usepackage{pifont}%
\newcommand{\xmark}{\textcolor{NegativeRed}{\ding{55}}\xspace}
\newcommand{\cmark}{\textcolor{ForestGreen}{\ding{51}}\xspace}

\makeatletter
\fancypagestyle{noconference}{%
  \fancyhf{}%
  \renewcommand{\headrulewidth}{\z@}%
  \renewcommand{\footrulewidth}{\z@}%
  \fancyfoot[C]{\if@ACM@printfolios\footnotesize\thepage\fi}%
  \fancyhead[LO]{\ACM@linecountL\@headfootfont\shorttitle}%
  \fancyhead[RE]{\@headfootfont\@shortauthors\ACM@linecountR}%
  \fancyhead[LE]{\ACM@linecountL}%
  \fancyhead[RO]{\ACM@linecountR}%
}
\AtBeginDocument{\pagestyle{noconference}}
\makeatother

\begin{document}

\title[LearnAct: Few-Shot Mobile GUI Agent with a Unified Demonstration Benchmark]{LearnAct: Few-Shot Mobile GUI Agent with a Unified Demonstration Benchmark}

\author{Guangyi Liu$^{\dag}$}
\affiliation{%
  \institution{Zhejiang University}
  \city{Hangzhou}
  \country{China}}

\author{Pengxiang Zhao$^{\dag}$}
\affiliation{%
  \institution{Zhejiang University}
  \city{Hangzhou}
  \country{China}}

\author{Liang Liu$^\ddag$}
\affiliation{%
  \institution{vivo AI Lab}
  \city{Hangzhou}
  \country{China}}

\author{Zhiming Chen}
\affiliation{%
  \institution{vivo AI Lab}
  \city{Hangzhou}
  \country{China}}

\author{Yuxiang Chai}
\affiliation{%
  \institution{vivo AI Lab}
  \city{Hangzhou}
  \country{China}}

\author{Shuai Ren}
\affiliation{%
  \institution{vivo AI Lab}
  \city{ShenZhen}
  \country{China}}

\author{Hao Wang}
\affiliation{%
  \institution{vivo AI Lab}
  \city{ShenZhen}
  \country{China}}

\author{Shibo He}
\affiliation{%
  \institution{Zhejiang University}
  \city{Hangzhou}
  \country{China}}

\author{Wenchao Meng\textsuperscript{\Letter}}
\affiliation{%
  \institution{Zhejiang University}
  \city{Hangzhou}
  \country{China}}
\email{wmengzju@zju.edu.cn}

\thanks{$^\dag$~Equal Contribution, \quad $^\ddag$~Project Lead,  \quad \Letter~
    Corresponding Author}

\renewcommand{\shortauthors}{Liu et al.}

\begin{abstract}
Mobile GUI agents show promise in automating tasks but face generalization challenges in diverse real-world scenarios. Traditional approaches using pre-training or fine-tuning with massive datasets struggle with the diversity of mobile applications and user-specific tasks. We propose enhancing mobile GUI agent capabilities through human demonstrations, focusing on improving performance in unseen scenarios rather than pursuing universal generalization through larger datasets.
To realize this paradigm, we introduce LearnGUI, the first comprehensive dataset specifically designed for studying demonstration-based learning in mobile GUI agents. It comprises 2,252 offline tasks and 101 online tasks with high-quality human demonstrations. We further develop LearnAct, a sophisticated multi-agent framework that automatically extracts knowledge from demonstrations to enhance task completion. This framework integrates three specialized agents: DemoParser for knowledge extraction, KnowSeeker for relevant knowledge retrieval, and ActExecutor for demonstration-enhanced task execution. 
Our experimental results show significant performance gains in both offline and online evaluations. In offline assessments, a single demonstration improves model performance, increasing Gemini-1.5-Pro's accuracy from 19.3\% to 51.7\%. In online evaluations, our framework enhances UI-TARS-7B-SFT's task success rate from 18.1\% to 32.8\%. LearnAct framework and LearnGUI benchmark establish demonstration-based learning as a promising direction for more adaptable, personalized, and deployable mobile GUI agents.
The project resources are available at \textcolor{blue}{\url{https://lgy0404.github.io/LearnAct}}.
\end{abstract}
\begin{teaserfigure}
	\centering
	\includegraphics[width=0.99\textwidth]{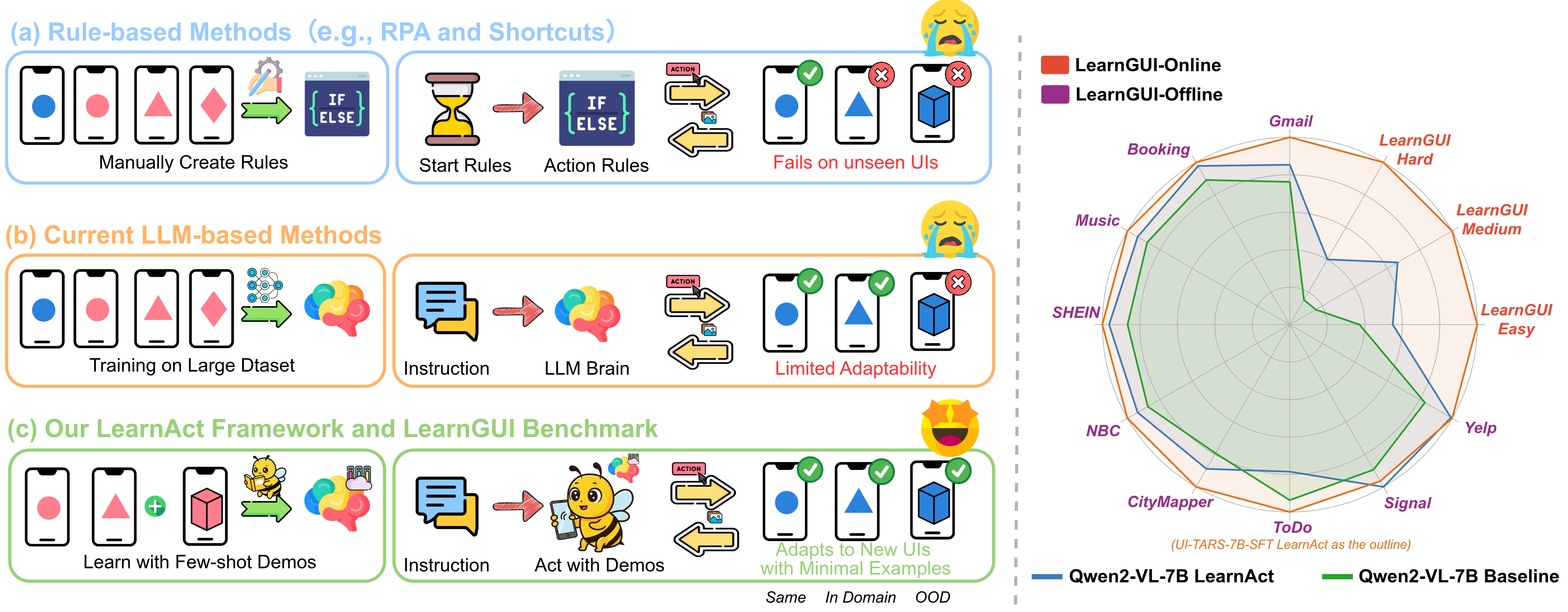}
        % \vspace{-8mm}
	\caption{
		\textbf{The LearnAct Framework and LearnGUI Benchmark focus on addressing the long-tail challenges in mobile GUI agent performance through demonstration-based learning.} 
	\textmd{From rule-based automation to LLM-powered agents, mobile GUI automation has evolved significantly, yet still struggles with long-tail scenarios due to interface diversity. Our LearnAct framework introduces demonstration-based learning to effectively handle these challenges, outperforming existing methods in both offline and online evaluations.}
	}
	\label{fig:overall}
\end{teaserfigure}

\maketitle
\section{Introduction}

\begin{figure}[t]
	\centering
	\includegraphics[width=0.47\textwidth]{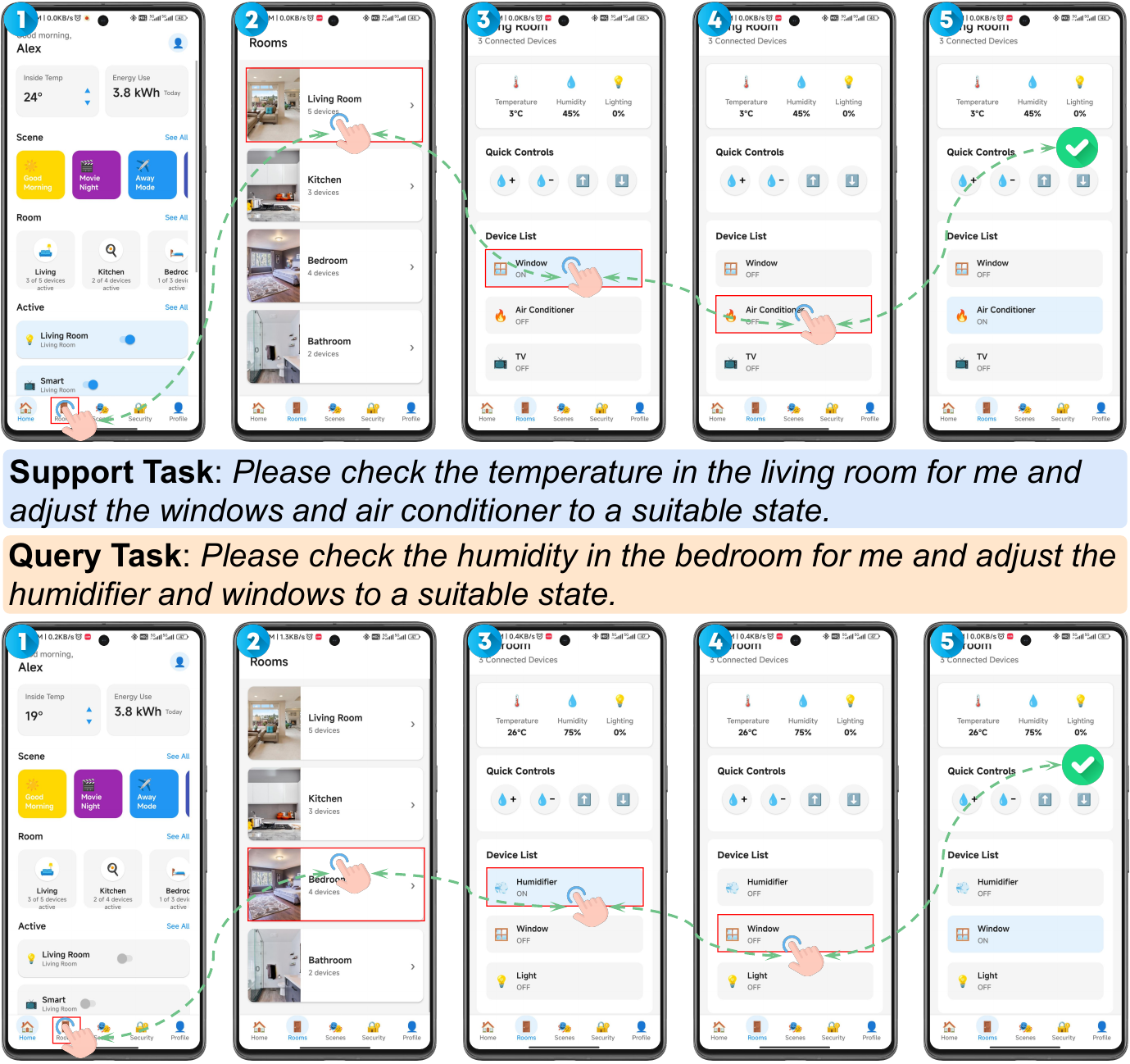}
	\caption{
		\textbf{A toy example for demonstration learning on mobile GUI Agent.}
	\textmd{We build a benchmark named LearnGUI for demonstration learning on Mobile GUI Agent, which provides different few-shot task combinations and offers multi-dimensional metrics including task similarity, UI similarity, and action similarity between support tasks and query tasks.}
	}
	\label{fig:learngui-data-example}
\end{figure}

Mobile device automation has evolved significantly over time, from simple rule-based scripts to sophisticated AI-powered agents~\cite{wu2024foundations,wang2024gui,zhang2024large,liu2025llm}. Traditional automation approaches like Robotic Process Automation (RPA)~\cite{agostinelli2019research} and rule-based shortcuts~\cite{guerreiro2008mnemonical,kennedy2011use} relied on predefined scripts to execute repetitive tasks, but they struggled with dynamic interfaces, required frequent maintenance when apps updated, and lacked understanding of complex user intentions.

More recently, mobile Graphical User Interface (GUI) agents have emerged as a transformative technology with the potential to revolutionize how humans interact with mobile devices. These agents leverage Large Language Models (LLMs) to autonomously complete human tasks through environmental interaction~\cite{wen2023droidbot,wen2024autodroid,wang2024mobileagentv1,wang2024mobileagentv2,liu2024vision,zhang2024mobileexperts,lu2024omniparser,chai2025a3,wang2025fedmobileagent}. They perceive phone states of mobile phone by observing screens (through screenshots or application UI trees) and generate actions (such as CLICK, TYPE, SWIPE, PRESS\_BACK, PRESS\_HOME, and PRESS\_ENTER) that are executed via the phone user interface~\cite{wu2024foundations,wang2024gui,zhang2024large,liu2025llm}. By harnessing the powerful perception and reasoning capabilities of LLMs, mobile GUI agents have the potential to fundamentally change how people interact with their mobile devices, bringing to life the "J.A.R.V.I.S." effect seen in science fiction.

Despite these promising advances, mobile GUI agents continue to face significant challenges in real-world deployment scenarios. The immense diversity of mobile applications and user interfaces creates pervasive long-tail scenarios where current agents struggle to perform effectively. The prevailing approaches to building modern mobile GUI agents rely on either the inherent capabilities of general-purpose LLMs~\cite{wen2023droidbot,wen2024autodroid,wang2024mobileagentv1,wang2024mobileagentv2,liu2024vision,zhang2024mobileexperts,lu2024omniparser,wang2025mobile} or fine-tuning with large volumes of data~\cite{zhang2023youautoui,hong2024cogagent,lin2024showui,xu2024aguvis}.  However, these methods face fundamental limitations when confronted with diverse real-world usage scenarios. As of 2025, billions of users interact with 1.68 million applications on Google Play alone~\cite{liu2025llm}, each with unique task requirements and UI layouts~\cite{wang2024gui,zhang2024large}. \textbf{Pre-training or fine-tuning datasets cannot feasibly cover this immense variety, leading to poor performance in unseen scenarios and hindering the widespread adoption of mobile GUI agents~\cite{li2024androidcontrol}}, as illustrated in Figure ~\ref{fig:overall} (left side). Traditional approaches simply cannot cover the entire spectrum of possible interactions and user-specific requirements across this heterogeneous landscape.

To address these limitations, we propose a novel paradigm that \textbf{enhances mobile GUI agent capabilities through few-shot demonstration learning}. Unlike traditional approaches that either lack flexibility or require massive datasets, our demonstration-based approach achieves both robustness and personalization by learning from a small number of user-provided examples. We recognize that mobile users have unique, repetitive tasks with inherent variability—such as smart home control with dynamic configurations, health monitoring with personalized parameters, or enterprise software with company-specific layouts. These scenarios combine stable patterns with variable elements, creating a "personalization gap" that pre-trained models cannot bridge. By leveraging user-specific demonstrations, our approach enables personalized assistants that learn both consistent patterns and adaptation strategies, acquiring task-specific knowledge impossible to cover in general training datasets. This personalization allows mobile GUI agents to overcome performance bottlenecks and provide truly helpful automation for the tasks users most want to delegate.

To fill the gap in high-quality demonstration data, we introduce \textbf{LearnGUI, the first dataset specifically designed to research and evaluate mobile GUI agents' ability to learn from few-shot demonstrations}. Built upon AMEX~\cite{chai2024amex} and AndroidWorld ~\cite{rawles2024androidworld}, LearnGUI comprises 2,252 offline few-shot tasks and 101 online tasks with high-quality human demonstrations. This dataset enables systematic research into demonstration-based learning for mobile GUI agents.  A toy example for LearnGUI is shown in Figure ~\ref{fig:learngui-data-example}.

Furthermore, we present \textbf{LearnAct, a multi-agent framework that automatically understands human demonstrations, generates instructional knowledge, and uses this knowledge to assist mobile GUI agents in reasoning about unseen scenarios}. LearnAct consists of three specialized agents: (1) DemoParser, a knowledge generation agent that extracts usable knowledge from demonstration trajectories to form a knowledge base; (2) KnowSeeker, a knowledge retrieval agent that searches the knowledge base for demonstration knowledge relevant to the current task; and (3) ActExecutor, a task execution agent that combines user instructions, real-time GUI environment, and retrieved demonstration knowledge to perform tasks effectively.

Our experimental results decisively validate the effectiveness of demonstration-based learning for mobile GUI agents, as shown in Figure ~\ref{fig:overall} (right side). In offline evaluations, a single demonstration dramatically improves model performance across diverse scenarios, with the most striking results seen in Gemini-1.5-Pro~\cite{team2024gemini}, whose accuracy increases from 19.3\% to 51.7\% (a 198.9\% relative improvement). Performance gains are particularly pronounced in complex applications, with accuracy in CityMapper increasing from 14.1\% to 69.4\% and in To-Do apps from 17.4\% to 69.2\%. For real-world online evaluations, our framework demonstrates exceptional effectiveness, with Qwen2-VL-7B~\cite{Qwen2VL} with LearnAct achieving significant performance gains, while UI-TARS-7B-SFT~\cite{qin2025ui}'s task success rate improves from 18.1\% to 32.8\% (+14.7\%). These findings offer a practical pathway to developing more adaptable and personalized mobile GUI agents.

In summary, our contributions are as follows:
\begin{itemize}    
    \item We develop LearnGUI, the first dataset designed for studying demonstration-based learning in mobile GUI agents, comprising 2,252 offline and 101 online tasks with high-quality human demonstrations.
    
    \item We design and implement LearnAct, a sophisticated multi-agent framework that systematically extracts, retrieves, and leverages knowledge from human demonstrations. This framework includes three specialized components: DemoParser (knowledge extraction), KnowSeeker (knowledge retrieval), and ActExecutor (task execution).
    
    \item Our evaluations demonstrate unprecedented performance gains: a single demonstration improves Gemini-1.5-Pro~\cite{team2024gemini}'s accuracy by 198.9\% in offline tests, while enhancing UI-TARS-7B-SFT~\cite{qin2025ui}'s online task success rate from 18.1\% to 32.8\%, advancing mobile GUI agents toward greater adaptability and practical deployability.
\end{itemize}
\section{Related Work}

\begin{table}[!t]
    \centering
    \footnotesize
    \renewcommand\arraystretch{1.1}
    \setlength{\tabcolsep}{2.7pt}
    \caption{\textbf{Comparison of different datasets and environments for benchmarking Mobile GUI agents.} \textmd{Column definitions: \# Inst. (number of instructions), \# Apps (number of applications), \# Step (average steps per task), Env. (supports environment interactions), HL (has high-level instructions), LL (has low-level instructions), GT (provides ground truth trajectories), FS (supports few-shot learning).}}
    \begin{tabular}{p{2.2cm}ccc>{\centering\arraybackslash}p{0.8cm}>{\centering\arraybackslash}p{0.45cm}>{\centering\arraybackslash}p{0.45cm}>{\centering\arraybackslash}p{0.45cm}>{\centering\arraybackslash}p{0.45cm}}
    \toprule
    \textbf{Dataset} & \textbf{\# Inst.} & \textbf{\# Apps} & \textbf{\# Step} & \textbf{Env.} & \textbf{HL} & \textbf{LL} & \textbf{GT} & \textbf{FS} \\
    \midrule
    %Rico~\cite{deka2017rico} & - & 9,772 & 15 & \xmark & \xmark & \xmark & \xmark & \xmark \\
    PixelHelp~\cite{li2020PixelHelp} & 187 & 4 & 4.2 & \xmark & \cmark & \xmark & \cmark & \xmark \\
    MoTIF~\cite{burns2021motif} & 276 & 125 & 4.5 & \xmark & \cmark & \cmark & \cmark & \xmark \\
    UIBert~\cite{bai2021uibert} & 16,660 & - & 1 & \xmark & \xmark & \cmark & \cmark & \xmark \\
    %Meta-GUI~\cite{sun2022metagui} & 1,125 & 11 & 15 & \xmark & \cmark & \xmark & \xmark & \xmark \\
    UGIF~\cite{venkatesh2022ugif} & 523 & 12 & 6.3 & \xmark & \cmark & \cmark & \cmark & \xmark \\
    AITW~\cite{rawles2024androidinthewild} & 30,378 & 357 & 6.5 & \xmark & \cmark & \xmark & \cmark & \xmark \\
    AITZ~\cite{zhang2024aitz} & 2,504 & 70 & 7.5 & \xmark & \cmark & \cmark & \cmark & \xmark \\
    %GUI Odyssey~\cite{lu2024guiodyssey} & 7,735 & 201 & 15 & \xmark & \cmark & \xmark & \xmark & \xmark \\
    AndroidControl~\cite{li2024androidcontrol} & 15,283 & 833 & 4.8 & \xmark & \cmark & \cmark & \cmark & \xmark \\
    AMEX~\cite{chai2024amex} & 2,946 & 110 & 12.8 & \xmark & \cmark & \xmark & \cmark & \xmark \\
    %MobileViews~\cite{gao2024mobileviews} & - & 21,053 & 15 & \xmark & \cmark & \xmark & \xmark & \xmark \\
    MobileAgentBench~\cite{wang2024mobileagentbench} & 100 & 10 & - & \xmark & \cmark & \xmark & \xmark & \xmark \\
    AppAgent~\cite{zhang2023appagent} & 50 & 10 & - & \xmark & \cmark & \xmark & \xmark & \xmark \\
    \midrule
    LlamaTouch~\cite{zhang2024llamatouch} & 496 & 57 & 7.01 & \cmark & \cmark & \xmark & \cmark & \xmark \\
    AndroidWorld~\cite{rawles2024androidworld} & 116 & 20 & - & \cmark & \cmark & \xmark & \xmark & \xmark \\
    AndroidLab~\cite{xu2024androidlab} & 138 & 9 & 8.5 & \cmark & \cmark & \xmark & \xmark & \xmark \\
    \midrule
    \textbf{LearnGUI (Ours)} & \textbf{2,353} & \textbf{73} & \textbf{13.2} & \textbf{\cmark} & \textbf{\cmark} & \textbf{\cmark} & \textbf{\cmark} & \textbf{\cmark} \\
    \bottomrule
    \end{tabular}
    \label{tab:datasets}
\end{table} 

\textbf{Mobile GUI Datasets and  Environments.} The development of mobile GUI agents relies heavily on high-quality datasets for training and evaluation. Table~\ref{tab:datasets} compares LearnGUI and existing mobile GUI datasets and benchmarks. These resources can be broadly categorized into static datasets and dynamic benchmarking environments.
Static datasets~\cite{li2020PixelHelp,burns2021motif,bai2021uibert,venkatesh2022ugif,rawles2024androidinthewild,zhang2024aitz,li2024androidcontrol,chai2024amex,wang2024mobileagentbench} typically provide natural language task descriptions, UI states (screenshots and/or application UI trees), and corresponding user actions (CLICK, SWIPE, TYPE, and other standardized interactions). These datasets vary in scale, ranging from hundreds to tens of thousands of instructions across different applications. Recent work like AppAgent~\cite{zhang2023appagent} has explored demonstration-based learning but without ground truth annotations or systematic analysis, providing only 50 tasks across 10 applications with high-level instructions. Notably, the average task length varies significantly across datasets, with AMEX~\cite{chai2024amex} featuring substantially longer sequences (12.8 steps on average) compared to AndroidControl (4.8 steps) and AITW~\cite{rawles2024androidinthewild} (6.5 steps).
Benchmarking environments, on the other hand, typically select a limited number of tasks and applications to provide dynamic testing environments~\cite{liu2025llm}. These frameworks evaluate agent performance through metrics such as task completion rates, critical state achievements, and execution time. Examples include LlamaTouch~\cite{zhang2024llamatouch}, AndroidWorld~\cite{rawles2024androidworld}, and AndroidLab~\cite{xu2024androidlab}, which offer interactive environments but lack few-shot demonstration capabilities.
We present the first systematic study of demonstration-based learning for mobile GUI agents through LearnGUI, which distinguishes itself through three key innovations. First, it is designed to evaluate few-shot learning capabilities with a comprehensive collection of 2,252 offline tasks and 101 online tasks. Built upon AMEX~\cite{chai2024amex} and AndroidWorld~\cite{rawles2024androidworld}, which feature longer, more complex tasks ideal for out-of-distribution and demonstration-based learning scenarios, LearnGUI provides a unified framework for both offline and online evaluation.
Second, while the original AMEX~\cite{chai2024amex} dataset contains 2,946 independent tasks unsuitable for few-shot evaluation, we conducted detailed analyses to transform and enhance this resource. Specifically, we made three key modifications: (1) Action Space Standardization, refining the original action space by removing inconsistent \texttt{TASK\_IMPOSSIBLE} actions, enhancing \texttt{TASK\_COMPLETE} to support information retrieval tasks, and standardizing formats for consistency; (2) K-shot Task Combinations, constructing systematic task groupings by recovering application context, computing instruction similarity within applications, and creating k-shot combinations with similar tasks as support demonstrations; and (3) Similarity Measurement, computing UI and action similarity through descriptive representations, enabling analysis of how different similarity types affect learning efficacy.
Third, regarding online evaluation, AndroidWorld~\cite{rawles2024androidworld} originally provides 116 dynamically constructed tasks without human demonstration trajectories. We collected 101 high-quality human demonstrations based on AndroidWorld's environment and dynamic instructions, forming LearnGUI-Online for evaluating the few-shot capabilities of mobile GUI agents in real-time scenarios.
By addressing the limitations of existing datasets, LearnGUI enables systematic research into few-shot learning for mobile GUI agents with varying k-shot configurations and controlled similarity conditions between support and query tasks.

\textbf{Mobile GUI Agents.} Mobile GUI agents are intelligent systems that leverage large language models to understand, plan, and execute tasks on mobile devices by integrating natural language processing, multimodal perception, and action execution capabilities~\cite{wu2024foundations,wang2024gui}. Recent developments in this field have explored various approaches to enhance agent performance and generalizability.
One prominent category of work focuses on designing effective prompting strategies to guide pre-trained LLMs without additional training~\cite{wei2022chain, yao2024tree, chen2022program}. By crafting prompts that incorporate task descriptions, interface states, and action histories, researchers can direct model behavior toward specific automation goals~\cite{wen2023droidbot, song2023navigating, wang2024mobileagentv1, wang2024mobileagentv2}. These approaches leverage the inherent capabilities of general-purpose LLMs but often struggle with complex tasks.
A second category involves adapting LLMs specifically for mobile automation through fine-tuning techniques~\cite{cheng2024seeclick, chen2024guicourse, lu2024guiodyssey, pawlowski2024tinyclick, hong2024cogagent, lin2024showui, xu2024aguvis}. These methods train models on GUI-specific data to enhance their understanding of and interaction with graphical interfaces. While improving performance over pre-training approaches, these fine-tuned models require substantial training data and still face generalization challenges.
Despite the progress made by both approaches, a fundamental limitation persists: the inability to generalize effectively to out-of-distribution scenarios. These methods both struggle with unseen applications, novel UI layouts, or unexpected task variations. These limitations stem from the impossibility of covering all potential real-world scenarios during training, creating significant bottlenecks in mobile GUI agent development.
To address these critical challenges, we introduce LearnAct, a sophisticated multi-agent framework that learns and reasons from screenshots without requiring UI tree information. The framework extracts, retrieves, and utilizes demonstration knowledge through three specialized components, enabling effective adaptation to new scenarios with minimal demonstrations.
\section{LearnGUI Dataset}
\subsection{Task Definition}

Mobile GUI tasks require agents to interact with digital environments by executing actions to fulfill user instructions. These tasks can be formally described as a \textbf{Partially Observable Markov Decision Process (POMDP)}, defined as $\mathcal{M} = (\mathcal{S}, \mathcal{O}, \mathcal{A}, \mathcal{T}, \mathcal{R})$, where $\mathcal{S}$ is the state space (current state of the mobile device), $\mathcal{O}$ is the observation space (instructions, screenshots, UI trees, etc.), $\mathcal{A}$ is the action space (e.g., click, type, swipe), $\mathcal{T}: \mathcal{S} \times \mathcal{A} \rightarrow \mathcal{S}$ is the state transition function, and $\mathcal{R}: \mathcal{S} \times \mathcal{A} \rightarrow [0,1]$ is the reward function. For example, a user might request the agent to "find the cheapest hotel in Paris for next weekend." The agent must perceive the current screen—either through an image or a UI tree—and execute a sequence of actions to complete the given task.

The key innovation in our approach is the integration of human demonstration knowledge into this POMDP framework. By incorporating demonstration knowledge $\mathcal{D}$ into the decision process, we enhance the agent's ability to handle out-of-distribution scenarios. This knowledge influences the agent's policy $\pi: \mathcal{O} \times \mathcal{D} \rightarrow \mathcal{A}$, which maps observations and relevant demonstration knowledge to actions, providing valuable examples of successful interaction patterns.

To study the impact of demonstration-based learning on mobile GUI agents, we need a dataset that provides various k-shot demonstrations with controlled similarity relationships between support and query tasks. This allows us to systematically investigate how demonstration quantity and task similarity affect agent performance. While cross-application knowledge transfer remains an interesting research direction, we focus on within-application task learning, as this represents the most practical use case where users would provide demonstrations for applications they frequently use.

Our dataset design specifically enables research on three key dimensions:
\begin{enumerate}
    \item \textbf{Unified comprehensive evaluation framework}: LearnGUI provides a standardized platform for studying few-shot demonstration learning in mobile GUI agents, featuring a unified action space and evaluation protocols that reflect real-world use cases
    \item \textbf{K-shot demonstration learning}: The dataset systematically explores how varying quantities of demonstrations (k=1, 2, or 3) affect agent performance, enabling research on the optimal number of examples needed
    \item \textbf{Multi-dimensional similarity analysis}: LearnGUI enables investigation of how different types of similarity between demonstration and query tasks influence learning efficacy and generalization capabilities
\end{enumerate}

This comprehensive approach allows for a nuanced analysis of how mobile GUI agents can leverage human demonstrations to improve task performance, especially in scenarios not covered by their training data.

\subsection{Data Collection}

The LearnGUI dataset consists of two components: \textbf{LearnGUI-Offline} for systematic evaluation of few-shot learning capabilities across varying similarity conditions, and \textbf{LearnGUI-Online} for real-time assessment in an interactive environment. Both components share a unified action space to ensure consistent evaluation, as detailed in Table~\ref{tab:action-space}.
\begin{table}[h]
    \centering
    \caption{\textbf{LearnGUI Action Space}}
    \begin{tabular}{>{\raggedright\arraybackslash}p{3.35cm} >{\raggedright\arraybackslash}p{4.5cm}}
    \toprule
    \textbf{Action} & \textbf{Definition} \\
    \midrule
    CLICK[x, y] & Click at coordinates (x, y). \\
    \midrule
    TYPE[text] & Type the specified text. \\
    \midrule
    SWIPE [direction] & Swipe in the specified direction. \\
    \midrule
    PRESS\_HOME & Go to the home screen. \\
    \midrule
    PRESS\_BACK & Go back to the previous app screen. \\
    \midrule
    PRESS\_ENTER & Press the enter button. \\
    \midrule
    TASK\_COMPLETE[answer] & Mark the task as complete. Provide answer inside brackets if required. \\
    \bottomrule
    \end{tabular}
    \label{tab:action-space}
\end{table} 

\subsubsection{LearnGUI-Offline}
We built LearnGUI-Offline by restructuring and enhancing the AMEX dataset~\cite{chai2024amex}, which contains 2,946 independent mobile tasks. To transform this resource for few-shot learning evaluation, we made several key modifications:

\textbf{Action Space Standardization.} We refined the original action space to better align with real-world scenarios. First, we removed \texttt{TASK\_IMPOSSIBLE} actions due to inconsistent labeling in the original dataset, which included errors such as tasks being incorrectly marked as impossible. Second, we enhanced \texttt{TASK\_COMPLETE} to \texttt{TASK\_COMPLETE[answer]} for information retrieval tasks. Many mobile tasks require returning specific information rather than just completion status. This aligns with both AMEX~\cite{chai2024amex} and AndroidWorld~\cite{rawles2024androidworld} paradigms.

\textbf{K-shot Task Combinations.} We constructed systematic k-shot task combinations through a multi-step process. We began by recovering the application context for each task through instruction and screenshot analysis, as the original dataset lacked explicit app labels. Next, we computed instruction similarity between tasks within the same application using the all-MiniLM-L6-v2 model. Finally, we created k-shot combinations (k=1,2,3) for each query task by selecting the k most similar tasks within the same application as support demonstrations, ensuring that the average similarity exceeded a minimum threshold of 0.6. This process yielded 2,252 tasks with valid k-shot combinations.

\textbf{Similarity Measurement.} To enable multi-dimensional similarity analysis, we computed metrics across three key dimensions. For Instruction Similarity, we utilized the scores calculated during the K-shot Task Combinations process. For UI Similarity, we merged the UI trees from all steps of each task and calculated similarity using TF-IDF vectorization and cosine similarity, capturing the visual and structural similarity of interfaces. For Action Similarity, following the DemoParser approach detailed in Section~\ref{sec:demoparser}, we generated descriptive representations of each action and computed embedding-based cosine similarity between task pairs.

\subsubsection{LearnGUI-Online}
For evaluating mobile GUI agents in real-time interactive scenarios, we developed LearnGUI-Online based on the AndroidWorld environment~\cite{rawles2024androidworld}. While AndroidWorld provides 116 dynamically constructed task templates, it lacks human demonstration trajectories essential for few-shot learning evaluation.

We identified 101 tasks suitable for human completion, excluding 15 tasks that proved challenging for human users. We then collected high-quality human demonstrations for these tasks. For tasks with dynamic elements, we generated specific instances and recorded corresponding demonstrations.

The resulting LearnGUI-Online dataset provides a realistic testbed for evaluating few-shot learning capabilities in mobile GUI agents under authentic conditions.

\begin{table*}[!t]
    \centering
    \caption{\textbf{Statistics of LearnGUI dataset splits.} \textmd{Each split is analyzed across multiple dimensions: Tasks (number of tasks), Apps (number of applications covered), Step actions (total action steps), similarity metrics (Avg Ins/UI/Act\textsubscript{Sim}), and distribution across four similarity profiles categorized by high (SH) and low (SL) UI and action similarity.}}
    \label{tab:dataset-splits}
    \resizebox{\linewidth}{!}{
    \begin{tabular}{l|*{11}{c}}
    \toprule
        \textbf{Split} & \textbf{K-shot} & \textbf{Tasks} & \textbf{Apps} & \textbf{Step actions} & \textbf{Avg Ins\textsubscript{Sim}} & \textbf{Avg UI\textsubscript{Sim}} & \textbf{Avg Act\textsubscript{Sim}} & \textbf{UI\textsubscript{SH}Act\textsubscript{SH}} & \textbf{UI\textsubscript{SH}Act\textsubscript{SL}} & \textbf{UI\textsubscript{SL}Act\textsubscript{SH}} & \textbf{UI\textsubscript{SL}Act\textsubscript{SL}} \\ 
    \midrule
        Offline-Train & 1-shot & 2,001 & 44 & 26,184 & 0.845 & 0.901 & 0.858 & 364 & 400 & 403 & 834 \\
        Offline-Train & 2-shot & 2,001 & 44 & 26,184 & 0.818 & 0.898 & 0.845 & 216 & 360 & 358 & 1,067 \\
        Offline-Train & 3-shot & 2,001 & 44 & 26,184 & 0.798 & 0.895 & 0.836 & 152 & 346 & 310 & 1,193 \\
    \midrule
        Offline-Test & 1-shot & 251 & 9 & 3,469 & 0.798 & 0.868 & 0.867 & 37 & 49 & 56 & 109 \\
        Offline-Test & 2-shot & 251 & 9 & 3,469 & 0.767 & 0.855 & 0.853 & 15 & 42 & 55 & 139 \\
        Offline-Test & 3-shot & 251 & 9 & 3,469 & 0.745 & 0.847 & 0.847 & 10 & 36 & 49 & 156 \\
    \midrule
        Online-Test & 1-shot & 101 & 20 & 1,423 & - & - & - & - & - & - & - \\
    \bottomrule
    \end{tabular}
    }
    \vspace{-0.1in}
\end{table*}

\subsection{Dataset Statistics}
Table~\ref{tab:datasets} presents the comprehensive statistics of the LearnGUI dataset in comparison with existing datasets. With 2,353 instructions across 73 applications and an average of 13.2 steps per task, LearnGUI offers rich data for studying demonstration-based learning in mobile GUI agents. The dataset provides various k-shot combinations (k=1,2,3) for each task, along with multi-dimensional similarity metrics across instruction, UI, and action dimensions. This design enables systematic analysis of how different types and quantities of demonstrations affect learning outcomes. The similarity distribution reflects the natural variation in mobile tasks within applications, with a meaningful spread across similarity levels that allows for a detailed investigation of knowledge transfer under different conditions. A detailed visualization of these similarity distributions is provided in Appendix~\ref{app:dataset_visualizations}.

\subsection{Dataset Splits}
\begin{figure}[t]
	\centering
	\includegraphics[width=0.45\textwidth]{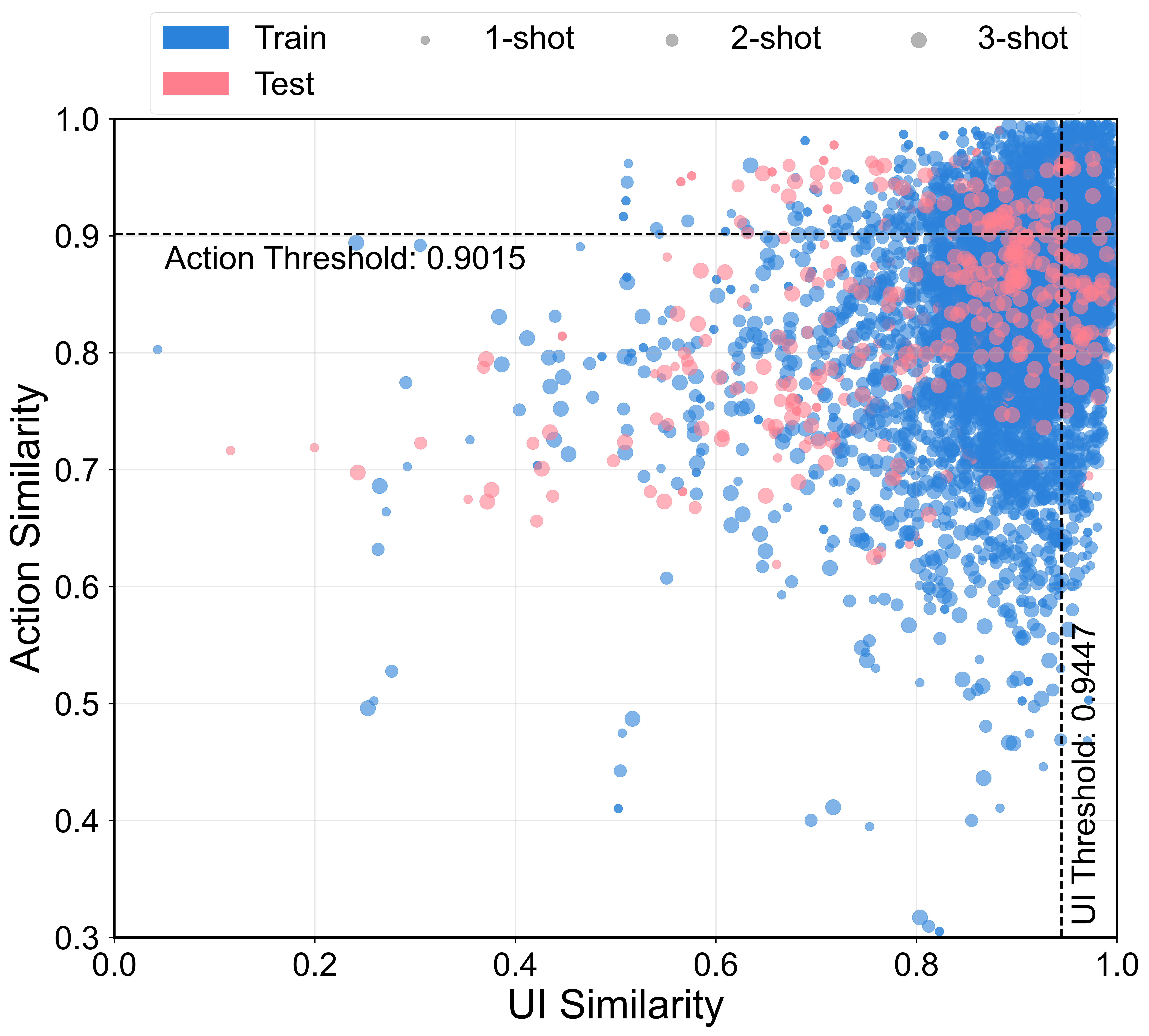}
	\caption{
		\textbf{Joint distribution of UI similarity and action similarity in LearnGUI-Offline.}
	\textmd{The scatter plot shows the relationship between UI and action similarity measures across task pairs. The quadrant divisions represent our categorization of tasks into four profiles: UI\textsubscript{SH}Act\textsubscript{SH}, UI\textsubscript{SH}Act\textsubscript{SL}, UI\textsubscript{SL}Act\textsubscript{SH}, and UI\textsubscript{SL}Act\textsubscript{SL}, enabling analysis of how different similarity combinations affect learning transfer.}
	}
	\label{fig:UI_Similarity_vs_Action_Similarity}
\end{figure}

We divided LearnGUI-Offline into training and testing splits to enable systematic evaluation of few-shot learning capabilities. Table~\ref{tab:dataset-splits} presents the detailed statistics of these splits, including the distribution of tasks across different similarity profiles.

The training set contains 2,001 tasks for each k-shot configuration (1, 2, and 3), spanning 44 applications with an average of 13.1 steps per task. The test set includes 251 tasks per k-shot configuration across 9 applications. Both splits maintain the same action space and similarity measurement methodology.

Based on empirical analysis, we established threshold values of 0.9447 for UI similarity and 0.9015 for action similarity to classify tasks into high (SH) and low (SL) similarity categories, enabling systematic analysis of how different similarity types affect learning from demonstrations.

As shown in Figure~\ref{fig:UI_Similarity_vs_Action_Similarity}, we classify tasks into four categories based on UI and action similarity:
\begin{itemize}
    \item \textbf{UI\textsubscript{SH}Act\textsubscript{SH}}: High UI similarity and high action similarity. For example, in a smart home app, two tasks that both involve adjusting the brightness of different lights in the living room would navigate through similar UI screens.
    \item \textbf{UI\textsubscript{SH}Act\textsubscript{SL}}: High UI similarity but low action similarity. For instance, in a smart home app, turning on all lights with a single button press versus adjusting each light's color temperature.
    \item \textbf{UI\textsubscript{SL}Act\textsubscript{SH}}: Low UI similarity but high action similarity. For example, setting a schedule for lights versus setting a schedule for the thermostat—different UI screens but similar action patterns.
    \item \textbf{UI\textsubscript{SL}Act\textsubscript{SL}}: Low UI similarity and low action similarity. For instance, checking security camera footage versus creating a scene that coordinates multiple devices.
\end{itemize}

This categorization enables a detailed analysis of how different types of similarity affect learning efficacy. For instance, we can investigate whether UI similarity or action similarity has a greater impact on successful knowledge transfer from demonstrations.

Additionally, the LearnGUI-Online test set contains 101 tasks across 20 applications. Unlike the offline dataset, these tasks are evaluated in real time through direct interaction with the mobile environment.

The comprehensive structure of LearnGUI, with its carefully designed splits and similarity classifications, provides a resource for studying how mobile GUI agents can learn from demonstrations under varying conditions of task similarity and demonstration quantity.
\section{Method: LearnAct}

\begin{figure*}[t]
	\centering
	\includegraphics[width=0.9\textwidth]{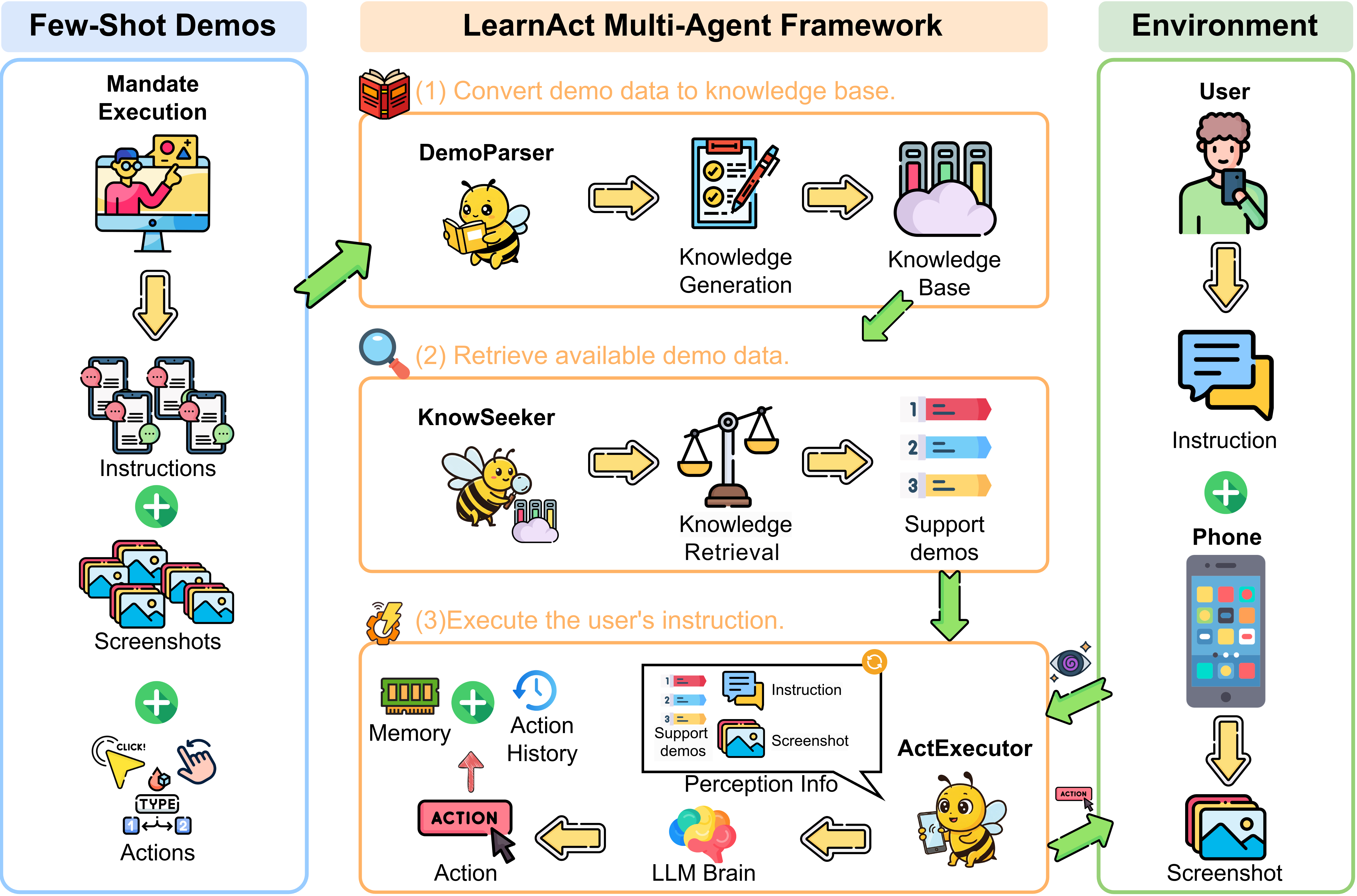}
	\caption{
		\textbf{Illustration of the overall framework of LearnAct.}
	\textmd{Architecture diagram showing the three main components (DemoParser, KnowSeeker, ActExecutor) and their interconnections within the LearnAct system, including data flow from human demonstrations to execution.}
	}
	\label{fig:learnact-pipline}
\end{figure*}

Building on the insights from our LearnGUI dataset, we introduce LearnAct, a novel framework designed to break through the limitations of traditional training approaches for mobile GUI agents. Rather than pursuing universal generalization through extensive training data, LearnAct establishes demonstration-based learning as a paradigm for developing more adaptable, personalized, and practically deployable mobile GUI agents. As illustrated in Figure~\ref{fig:learnact-pipline}, LearnAct is a sophisticated multi-agent framework that automatically understands human demonstrations, generates instructional knowledge, and leverages this knowledge to assist mobile GUI agents in reasoning about unseen scenarios.
The LearnAct framework consists of three specialized components, each addressing a critical aspect of demonstration-based learning: (1) DemoParser (Section~\ref{sec:demoparser}), a knowledge generation agent that extracts usable knowledge from demonstration trajectories to form a knowledge base; (2) KnowSeeker (Section~\ref{sec:knowseeker}), a knowledge retrieval agent that searches the knowledge base for demonstration knowledge relevant to the current task; and (3) ActExecutor (Section~\ref{sec:actexecutor}), a task execution agent that combines user instructions, real-time GUI environment, and retrieved demonstration knowledge to perform tasks effectively.

\subsection{DemoParser}
\label{sec:demoparser}
\begin{figure*}[t]
	\centering
	\includegraphics[width=0.9\textwidth]{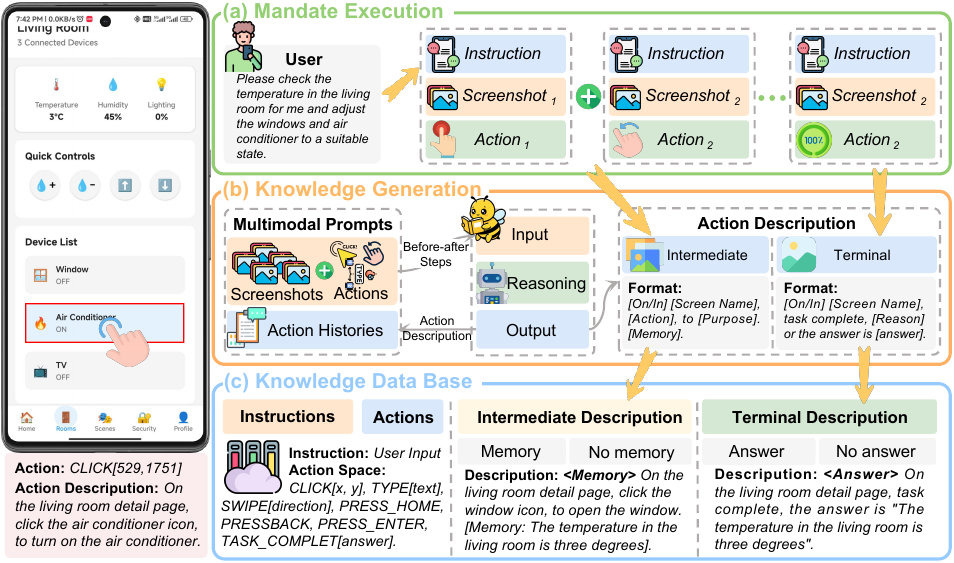}
	\caption{
		\textbf{Pipeline of DemoParser Agent.}
	\textmd{Input instructions and corresponding actions and screenshots; output low-level action descriptions and create knowledge database.
	This process transforms high-level user instructions into precise operation sequences while building a reusable domain knowledge base to improve mobile interface interaction automation efficiency.}
	}
	\label{fig:demoparser-pipline}
\end{figure*}

The DemoParser transforms raw human demonstrations into structured demonstration knowledge. It takes as input a raw action sequence (consisting of coordinates-based clicks, swipes, and text inputs) along with corresponding screenshots and task instructions. It then utilizes a vision-language model to generate semantically descriptive action descriptions that capture the essence of each demonstration step (e.g., ``On Search Page, click the search box, to enter keywords''). Building on these descriptions, it constructs a structured knowledge base that records both the high-level action semantics and the contexts in which they occur, as shown in Figure~\ref{fig:demoparser-pipline}.

Formally, DemoParser implements a knowledge generation function $G: \mathcal{I} \times \mathcal{S} \times \mathcal{A} \rightarrow \mathcal{K}$, where $\mathcal{I}$ represents the space of instructions, $\mathcal{S}$ is the space of screenshot sequences, $\mathcal{A}$ is the space of action sequences, and $\mathcal{K}$ is the knowledge space. For each demonstration trajectory $(i, s, a) \in \mathcal{I} \times \mathcal{S} \times \mathcal{A}$, DemoParser generates a knowledge entry $k \in \mathcal{K}$ that encapsulates the demonstration in a semantically descriptive format, converting raw coordinate-based actions (e.g., CLICK[123,456]) into meaningful operation descriptions (e.g., "click search box").

The knowledge generation process is decomposed into a sequence of description generation steps for each action in the demonstration trajectory. Let $d_j$ represent the description for action $a_j$, which is generated using a context-aware description function $\delta: \mathcal{I} \times \mathcal{A}_j \times \mathcal{V}_j \times \mathcal{H}_{j-1} \rightarrow \mathcal{D}$, where $\mathcal{V}_j$ is the visual representation of action $a_j$ execution and $\mathcal{H}_{j-1} = \{d_1, d_2, \ldots, d_{j-1}\}$ is the history of previous action descriptions.

Algorithm~\ref{alg:demoparser_appendix} in Appendix~\ref{app:algorithm_details} outlines the knowledge generation process. For each demonstration, DemoParser preserves the original task instruction and action sequence while generating semantically descriptive action descriptions. These descriptions provide crucial context about the purpose and significance of each action in the demonstration, enabling more effective knowledge transfer to new scenarios.

For intermediate actions, DemoParser analyzes a visual representation of the action execution, showing before-action and after-action screenshots with the action visualized (e.g., click locations highlighted). The framework combines this visual input with the task instruction, action history, and current action to generate a description that follows a standardized format: "[On/In] [Screen Name], [Action Details], to [Purpose]". For example: "On Home Screen, tap 'Settings' icon, to access device configuration."
For terminal actions, DemoParser processes the final screenshot, task instruction, and complete action history to generate a conclusion in the format: "[On/In] [Screen], complete task, [Reason/Answer]"

A distinctive feature of DemoParser is its memory mechanism, which captures critical information observed during task execution that may be necessary for future steps. The model identifies and annotates task-relevant information that is directly related to the user's instruction, will likely be needed in subsequent steps, and has not been previously recorded. These memory annotations are included in the action descriptions when appropriate: "[On/In] [Screen], [Action], to [Purpose]. [Memory: important information for future steps]". For example, in a shopping task, a memory annotation might capture: "[Memory: iPhone 13 Pro costs \$999 with 128GB storage]". The detailed prompt for this memory mechanism is provided in Appendix~\ref{app:demoparser_prompts}.

This memory mechanism is particularly valuable for complex tasks requiring information retention across multiple steps, such as comparing prices, remembering account details, or tracking status changes. By transforming raw demonstrations into structured, semantically descriptive knowledge with memory capabilities, DemoParser enables effective knowledge transfer from human demonstrations to automated task execution.

\subsection{KnowSeeker}
\label{sec:knowseeker}
\begin{figure}[t]
	\centering
	\includegraphics[width=0.49\textwidth]{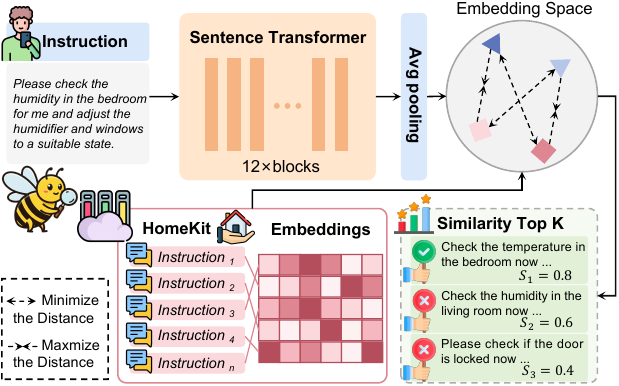}
	\caption{
		\textbf{Pipeline of KnowSeeker Agent.}
	\textmd{The KnowSeeker Agent converts demo trajectories from the knowledge base into a vector database. When executing user tasks, KnowSeeker retrieves the top-k relevant demos from the vector database for subsequent use. This approach enables efficient retrieval of similar demonstrations to assist with new task execution.}
	}
	\label{fig:knowseeker-pipline}
\end{figure}

KnowSeeker is the retrieval component of the LearnAct framework that identifies demonstration knowledge most relevant to the current task context. As depicted in Figure~\ref{fig:knowseeker-pipline}, this agent serves as the bridge between the knowledge base generated by DemoParser and the execution environment of ActExecutor. While DemoParser focuses on transforming demonstrations into structured knowledge, KnowSeeker specializes in efficiently accessing and selecting the most applicable knowledge for a specific task, addressing the critical challenge of knowledge relevance in few-shot learning scenarios.

Formally, KnowSeeker implements a retrieval function $R: \mathcal{I} \times \mathcal{K} \rightarrow \mathcal{K}^{(s)}$, where $\mathcal{I}$ is the instruction space, $\mathcal{K}$ is the knowledge base, and $\mathcal{K}^{(s)} \subset \mathcal{K}$ is a subset of knowledge entries determined to be relevant for the given instruction. This retrieval process is crucial for effective knowledge utilization, as it filters the potentially vast knowledge base to focus exclusively on demonstrations that offer valuable insights for the current task.

The core of KnowSeeker's retrieval mechanism relies on semantic similarity measurement between the current task instruction and the instructions associated with demonstrations in the knowledge base. This similarity-based retrieval can be formally defined as:

\begin{equation}
R(i, K) = \{k_j \in K \mid sim(i, i_j) \geq \tau_s \}_{j=1}^{top-k}
\end{equation}

where $i$ is the current instruction, $i_j$ is the instruction associated with knowledge entry $k_j$, $sim(\cdot, \cdot)$ is a similarity function, $\tau_s$ is a similarity threshold, and $top-k$ indicates selection of the $k$ most similar entries.

To implement this similarity measurement efficiently, KnowSeeker employs a two-phase approach:

\begin{enumerate}
    \item \textbf{Embedding Generation:} Instructions are transformed into dense vector representations using a pre-trained sentence transformer model. Specifically, we utilize the all-MiniLM-L6-v2 model, which offers an optimal balance between computational efficiency and semantic representational power. This model has been fine-tuned on diverse natural language understanding tasks, making it particularly well-suited for capturing the semantic essence of mobile GUI task instructions.
    
    \item \textbf{Similarity Computation:} The cosine similarity between embedding vectors is calculated to quantify the semantic relationship between instructions. For instructions $i$ and $i_j$ with corresponding embeddings $e_i$ and $e_j$, the similarity is computed as:
    
    \begin{equation}
    sim(i, i_j) = \frac{e_i \cdot e_j}{||e_i|| \cdot ||e_j||}
    \end{equation}
\end{enumerate}

To optimize retrieval efficiency, KnowSeeker pre-computes embeddings for all instructions in the knowledge base during initialization. This approach transforms the potentially expensive operation of computing pairwise similarities during runtime into a more manageable vector comparison task. The pre-computation process is described as:

\begin{equation}
E = \{e_j = f_{embed}(i_j) \mid k_j \in K\}
\end{equation}

where $f_{embed}$ is the embedding function implemented by the sentence transformer model.

During task execution, when presented with a new instruction $i$, KnowSeeker:
1. Computes the embedding $e_i = f_{embed}(i)$
2. Calculates similarities $S = \{sim(e_i, e_j) \mid e_j \in E\}$
3. Selects the top-$k$ knowledge entries based on similarity scores

This approach ensures that knowledge retrieval scales efficiently with the size of the knowledge base, enabling rapid identification of relevant demonstrations even as the framework's experiential knowledge grows over time. By systematically identifying the most relevant demonstration knowledge, KnowSeeker enables ActExecutor to perform tasks more effectively, particularly in unfamiliar scenarios.

\subsection{ActExecutor}
\label{sec:actexecutor}
\begin{figure*}[t]
	\centering
	\includegraphics[width=0.9\textwidth]{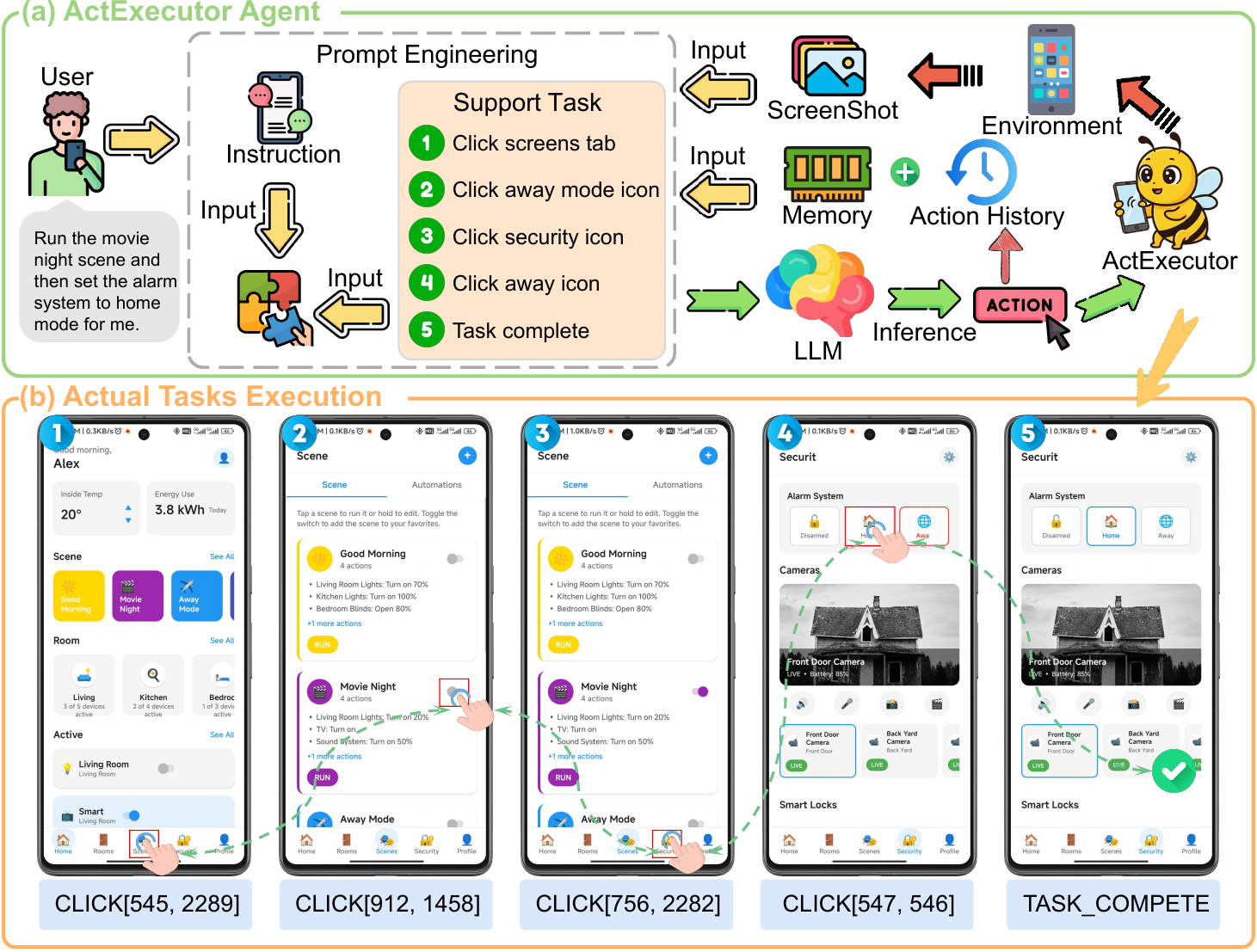}
	\caption{
		\textbf{Pipeline of ActExecutor Agent.}
	\textmd{The ActExecutor Agent executes the low-level action descriptions generated by the Action Planner Agent. It uses the KnowSeeker Agent to retrieve relevant demonstrations from the knowledge base and execute the actions in the demonstrations. This approach enables efficient execution of low-level actions to assist with new task execution.}
	}
	\label{fig:actexecutor-pipline}
\end{figure*}

ActExecutor is the execution component of the LearnAct framework that translates retrieved demonstration knowledge into effective actions in the target environment. As illustrated in Figure~\ref{fig:actexecutor-pipline}, this agent represents the culmination of the LearnAct pipeline, integrating user instructions, real-time GUI observations, and demonstration knowledge to navigate even unfamiliar mobile applications successfully. While DemoParser creates structured knowledge and KnowSeeker retrieves relevant demonstrations, ActExecutor applies this knowledge to solve practical tasks, addressing the critical challenge of knowledge utilization in few-shot learning scenarios.

ActExecutor implements the POMDP framework introduced earlier, with the critical enhancement of incorporating demonstration knowledge into the decision-making process. The execution process can be formally described as a sequential decision-making loop that iteratively selects actions $a_t \in \mathcal{A}$ based on current observations $o_t \in \mathcal{O}$ and demonstration knowledge $\mathcal{D}$, following policy $\pi: \mathcal{O} \times \mathcal{D} \rightarrow \mathcal{A}$.

The ActExecutor policy $\pi$ is implemented through a large vision-language model that processes a carefully constructed prompt integrating all available information sources. This prompt-based policy can be expressed as:

\begin{equation}
\pi(o_t, \mathcal{D}) = f_{LLM}(P(i, o_t, h_{t-1}, \mathcal{D}))
\end{equation}

where $i$ is the user instruction, $o_t$ is the current observation (screenshot), $h_{t-1}$ is the action history up to time $t-1$, $\mathcal{D}$ is the retrieved demonstration knowledge, $P$ is a prompt construction function, and $f_{LLM}$ is the LLM-based decision function.

Algorithm~\ref{alg:actexecutor_appendix} in Appendix~\ref{app:algorithm_details} outlines the execution process. For each task, ActExecutor processes the user instruction and screenshot observations through a sequence of perception, decision, and action phases until the task is completed or a maximum step limit is reached.

The execution process integrates three key phases:

\begin{enumerate}
    \item \textbf{Perception Phase:} ActExecutor perceives the current state of the mobile device through screenshot observations $o_t$. These observations provide the visual context essential for understanding the available interaction options and current application state.
    
    \item \textbf{Decision Phase:} The agent constructs a comprehensive prompt that integrates the user instruction $i$, current observation $o_t$, action history $h$, and retrieved demonstrations $\mathcal{D}$. This prompt is processed by a large vision-language model using templates detailed in Appendix~\ref{app:actexecutor_prompts}, resulting in a selected action from the predefined action space described in Table~\ref{tab:action-space}.
    
    \item \textbf{Action Phase:} The selected action $a_t$ is executed in the mobile environment, generating a state transition according to the transition function $\mathcal{T}$ of the POMDP. Additionally, the agent generates a description $d_t$ of the executed action using a process similar to DemoParser's description generation, which serves as part of the action history for subsequent steps.
\end{enumerate}

The prompt construction function $P$ plays a critical role in ActExecutor's effectiveness. It integrates the agent's role definition, demonstration examples, task and observation context, action history, and the action space definition into a comprehensive prompt that guides the model's decision-making.

This approach enables ActExecutor to leverage demonstrations as exemplars that guide its decision-making process. When faced with a novel UI state, the agent identifies analogous situations from demonstrations and adapts the demonstrated actions to the current context. This capability is particularly valuable for handling out-of-distribution scenarios where the agent lacks direct experience.

By closing the loop between demonstration knowledge and task execution, ActExecutor completes the LearnAct framework's end-to-end pipeline for demonstration-based learning. The combination of knowledge generation (DemoParser), knowledge retrieval (KnowSeeker), and knowledge-guided execution (ActExecutor) enables effective few-shot learning for mobile GUI agents, addressing the fundamental challenge of generalization to unseen scenarios with minimal examples.

\section{Experiments}

We conducted comprehensive evaluations of the LearnAct framework through both offline and online experiments. The offline experiments were performed on the LearnGUI-Offline dataset to evaluate step-by-step task execution capabilities, while the online experiments utilized the LearnGUI-Online platform to assess end-to-end task completion in real-world interactive scenarios. We evaluated a diverse set of models, including both commercial (e.g., Gemini-1.5-Pro~\cite{team2024gemini}) and open-source models (e.g., UI-TARS-7B-SFT~\cite{qin2025ui}, Qwen2-VL-7B~\cite{Qwen2VL}), to demonstrate the broad applicability of our approach across different model architectures and capabilities.

\subsection{Experiment Setup}

The diverse similarity profiles in LearnGUI provide a unique opportunity to evaluate mobile GUI agents' capabilities. Our experiments have two primary goals: (1) to evaluate the feasibility and effectiveness of enhancing mobile agents through few-shot demonstrations as a means to overcome the limitations of traditional pre-training or fine-tuning approaches; and (2) to investigate how different factors such as demonstration quantity (k=1,2,3) and various similarity aspects (instruction, UI, and action) influence the effectiveness of demonstration-based learning.

\textbf{Implementation Details.} We conducted experiments with three foundation models: Gemini-1.5-Pro~\cite{team2024gemini}, UI-TARS-7B-SFT~\cite{qin2025ui}, and Qwen2-VL-7B~\cite{Qwen2VL}. For all models, we set the temperature to zero to obtain deterministic responses. For Qwen2-VL-7B~\cite{Qwen2VL} and UI-TARS-7B-SFT~\cite{qin2025ui}, we employed parameter-efficient fine-tuning using LoRA with rank 64, alpha 128, and dropout probability 0.1. We targeted all modules while freezing the vision encoder to ensure computational efficiency. Training used a learning rate of 1e-5 with cosine scheduling, batch size of 1, gradient accumulation over 8 steps, a warmup ratio of 0.001, and was conducted for 1 epoch. All fine-tuning experiments were conducted on 8 NVIDIA L40S GPUs. For offline experiments, Gemini-1.5-Pro~\cite{team2024gemini} was evaluated directly on the LearnGUI-Offline test set without additional training. UI-TARS-7B-SFT~\cite{qin2025ui} and Qwen2-VL-7B~\cite{Qwen2VL} were fine-tuned on the LearnGUI-Offline training set before evaluation. For online experiments, we deployed all models except Gemini-1.5-Pro~\cite{team2024gemini} (which showed limited task completion capabilities in preliminary tests despite accuracy improvements) to the LearnGUI-Online environment, using 1-shot demonstration retrieval for all LearnAct-enhanced models.

\textbf{Baselines.} To rigorously evaluate our approach, we compared LearnAct against several baselines. These include: (1) SPHINX-GUI Agent, the original agent developed for the AMEX dataset~\cite{chai2024amex}, providing a reference point for task execution on similar data; (2) Zero-shot inference versions of all models (Gemini-1.5-Pro~\cite{team2024gemini}, UI-TARS-7B-SFT~\cite{qin2025ui}, and Qwen2-VL-7B~\cite{Qwen2VL}) within the LearnAct framework but without demonstration knowledge, maintaining identical execution environments for fair comparison; and (3) For online evaluation, we additionally compared against GPT-4o, Gemini-Pro-1.5, Claude Computer-Use, and Aguvis to benchmark against current advanced systems.

\textbf{Evaluation Metrics.} For offline evaluation, we adopted mainstream evaluation protocols widely used in recent mobile GUI agent research, such as UI-TARS~\cite{qin2025ui} and OS-ATLAS~\cite{wu2024atlas}. Specifically, we measured step accuracy, which consists of two components: action type accuracy and action match accuracy. Action type accuracy measures the percentage of steps where the predicted action type (CLICK, TYPE, SWIPE, etc.) matches the ground truth. Action match accuracy measures the percentage of steps where both the action type and its parameters are correct, following standard evaluation criteria. For CLICK actions, coordinates are considered correct if they fall within 14\% of the screen width from the ground truth. For TYPE actions, the content is correct if the F1 score between prediction and ground truth exceeds 0.5. For SWIPE actions, the direction must precisely match the ground truth. For other actions (e.g., PRESS\_BACK), an exact match is required. For TASK\_COMPLETE actions, we only verify the action type and ignore the answer field. For online evaluation, we measured the task success rate (SR), which represents the percentage of tasks completed successfully in the real-time interactive environment.

\subsection{Main Results}

\subsubsection{Offline Agent Capability Evaluation}

\begin{table*}[t]
	\definecolor{ForestGreen}{RGB}{34,139,34}
	\definecolor{NegativeRed}{RGB}{220,20,60}
	\caption{\textbf{Performance comparison of mobile GUI agents on LearnGUI-Offline dataset (action match accuracy \%).} \textmd{Results show absolute values and relative improvements [in brackets] compared to baselines. Performance is evaluated across different models and number of support examples (1/2/3-shot).}
	}
    \vspace{-2mm}
    \label{tab:amex_overall}
    \begin{center}
    \setlength{\tabcolsep}{3pt}
    \scalebox{0.99}
    {
    \begin{tabular}{l|c|c|c|*{9}{c}}
    \toprule
    \multicolumn{1}{c}{\bf Models} & \multicolumn{1}{c}{\bf Method} & \multicolumn{1}{c}{\bf Supports} & \multicolumn{1}{c}{\bf Average} & Gmail & Booking & Music & SHEIN & NBC & CityMapper & ToDo & Signal & Yelp
    \\
    \midrule
    SPHINX-GUI Agent\cite{chai2024amex} & AMEX & 0-shot & 67.2 & 45.9 & 64.5 & 74.4 & 71.8 & 70.3 & 67.4 & 79.3 & 64.9 & 66.3 \\
    \midrule
    \multirow{4}{*}{gemini-1.5-pro} & Baseline & 0-shot & 19.3 & 20.1 & 16.4 & 24.5 & 10.2 & 35.6 & 14.1 & 17.4 & 27.9 & 15.2\\
    & \multirow{3}{*}{LearnAct} & 1-shot & 51.7 \textcolor{ForestGreen}{[+32.4]} & 55.5 & 47.1 & \textbf{60.0} & 35.7 & \textbf{56.4} & 54.7 & 60.6 & 63.1 & 54.6\\
    & & 2-shot & \underline{55.6} \textcolor{ForestGreen}{[+36.3]} & \underline{57.5} & \underline{53.2} & \underline{55.3} & \underline{39.6} & \underline{56.1} & \underline{58.2} & \underline{68.1} & \underline{69.7} & \underline{60.0}\\
    & & 3-shot & \textbf{57.7} \textcolor{ForestGreen}{[+38.4]} & \textbf{58.4} & \textbf{56.6} & 54.6 & \textbf{43.9} & 53.9 & \textbf{69.4} & \textbf{69.2} & \textbf{70.5} & 57.6\\
    \midrule
    \multirow{4}{*}{UI-TARS-7B-SFT} & Baseline & 0-shot & 77.5 & 68.1 & 81.0 & 81.1 & 72.9 & 80.9 & 70.6 & 66.0 & \underline{92.6} & 82.4\\
    & \multirow{3}{*}{LearnAct} & 1-shot & \textbf{82.8} \textcolor{ForestGreen}{[+5.3]} & \underline{79.9} & \textbf{82.9} & \textbf{86.6} & \underline{75.7} & \underline{86.3} & \underline{79.4} & \underline{84.0} & \textbf{89.3} & \underline{83.0}\\
    & & 2-shot & \underline{81.9} \textcolor{ForestGreen}{[+4.4]} & \textbf{80.1} & \underline{80.7} & \underline{86.2} & \textbf{76.1} & \underline{87.2} & 80.0 & 83.7 & 84.4 & \textbf{84.2}\\
    & & 3-shot & 82.1 \textcolor{ForestGreen}{[+4.6]} & \underline{79.9} & 80.9 & \underline{86.2} & \underline{75.7} & \textbf{86.9} & \textbf{81.2} & \textbf{85.8} & 84.4 & \textbf{84.2}\\
    \midrule
    \multirow{4}{*}{Qwen2-VL-7B} & Baseline & 0-shot & 71.8 & 60.8 & 73.9 & 76.0 & 65.5 & 75.5 & 62.9 & 78.7 & 82.8 & 69.1\\
    & \multirow{3}{*}{LearnAct} & 1-shot & 77.3 \textcolor{ForestGreen}{[+5.5]} & \textbf{75.0} & \underline{77.5} & \underline{77.8} & \underline{69.8} & \underline{83.5} & \underline{72.9} & \underline{78.0} & \underline{83.6} & \underline{78.8}\\
    & & 2-shot & \underline{78.5} \textcolor{ForestGreen}{[+6.7]} & \textbf{75.0} & 78.0 & \underline{77.8} & \textbf{73.3} & \underline{86.0} & 73.5 & \underline{81.9} & \textbf{87.7} & 77.6\\
    & & 3-shot & \textbf{79.4} \textcolor{ForestGreen}{[+7.6]} & \textbf{75.0} & \textbf{78.8} & \textbf{78.6} & 72.6 & \textbf{87.8} & \textbf{77.1} & \textbf{82.6} & \textbf{87.7} & \textbf{80.6}\\
    \bottomrule
    \end{tabular}
    }
    \end{center}
    \end{table*} 

Table~\ref{tab:amex_overall} presents the performance comparison of different models on the LearnGUI-Offline dataset. The results demonstrate the substantial improvements achieved by the LearnAct framework across all tested models. Gemini-1.5-Pro~\cite{team2024gemini} shows the most dramatic improvement, with performance increasing from 19.3\% to 51.7\% (+32.4\%) with just a single demonstration, and further improving to 57.7\% (+38.4\%) with three demonstrations. This represents a 198.9\% relative improvement, highlighting the powerful potential of demonstration-based learning even for advanced foundation models. UI-TARS-7B-SFT~\cite{qin2025ui}, despite already having strong zero-shot performance (77.5\%), still achieves significant gains with LearnAct, reaching 82.8\% (+5.3\%) with a single demonstration. This indicates that even models specifically fine-tuned for GUI tasks can benefit from demonstration knowledge. Qwen2-VL-7B~\cite{Qwen2VL} demonstrates consistent improvement from 71.8\% to 77.3\% (+5.5\%) with one demonstration, and to 79.4\% (+7.6\%) with three demonstrations, confirming that the benefits of LearnAct generalize across models with different architectures and capabilities.

The results also reveal interesting patterns regarding the impact of demonstration quantity. For Gemini-1.5-Pro~\cite{team2024gemini}, performance scales monotonically with the number of demonstrations, suggesting that less specialized foundation models can benefit substantially from additional examples. In contrast, UI-TARS-7B-SFT~\cite{qin2025ui} achieves its peak performance with just one demonstration, indicating that models already fine-tuned for GUI tasks may efficiently extract necessary information from minimal demonstrations.

Application-specific results highlight LearnAct's consistent improvement across diverse scenarios, with particularly notable gains in complex applications like CityMapper (from 14.1\% to 69.4\% for Gemini-1.5-Pro~\cite{team2024gemini}) and To-Do apps (from 17.4\% to 69.2\%). This suggests that demonstration-based learning is especially valuable for navigating applications with complex interactions and non-standard interfaces.

\begin{table*}[t]
    \definecolor{ForestGreen}{RGB}{34,139,34}
    \caption{
        \textbf{Performance breakdown of LearnAct-Offline on different UI and action combinations.}
    \textmd{Performance metrics (type and match accuracy) across four similarity quadrants showing absolute values and relative improvements [in brackets] compared to baselines. Results are grouped by model and number of support examples (1/2/3-shot).}
    }
    \vspace{-2mm}
    \label{tab:breakdown}
    \begin{center}
    \setlength{\tabcolsep}{3pt}
    \scalebox{0.99}
    {
    \begin{tabular}{l|c|*{8}{c}}
    \toprule
    \multicolumn{1}{c}{\bf Models} & \multicolumn{1}{c}{\bf Supports} & \multicolumn{2}{c}{\bf UI\textsubscript{SH}Act\textsubscript{SH}} & \multicolumn{2}{c}{\bf UI\textsubscript{SH}Act\textsubscript{SL}} & \multicolumn{2}{c}{\bf UI\textsubscript{SL}Act\textsubscript{SH}} & \multicolumn{2}{c}{\bf UI\textsubscript{SL}Act\textsubscript{SL}} \\
    \cmidrule(lr){3-4} \cmidrule(lr){5-6} \cmidrule(lr){7-8} \cmidrule(lr){9-10}
    & & type & match & type & match & type & match & type & match \\
    \midrule
    \multirow{3}{*}{gemini-1.5-pro} & 1-shot & 79.5 \textcolor{ForestGreen}{[+12.8]} & 50.2 \textcolor{ForestGreen}{[+35.6]} & 78.1 \textcolor{ForestGreen}{[+12.3]} & 47.8 \textcolor{ForestGreen}{[+33.2]} & 77.5 \textcolor{ForestGreen}{[+9.2]} & 52.3 \textcolor{ForestGreen}{[+30.5]} & 77.9 \textcolor{ForestGreen}{[+14.1]} & 44.2 \textcolor{ForestGreen}{[+29.3]} \\
    & 2-shot & 77.7 \textcolor{ForestGreen}{[+13.0]} & 53.9 \textcolor{ForestGreen}{[+37.3]} & 73.2 \textcolor{ForestGreen}{[+10.8]} & 49.9 \textcolor{ForestGreen}{[+34.7]} & 80.0 \textcolor{ForestGreen}{[+9.0]} & 56.5 \textcolor{ForestGreen}{[+34.8]} & 77.2 \textcolor{ForestGreen}{[+12.9]} & 48.9 \textcolor{ForestGreen}{[+34.4]} \\
    & 3-shot & 72.3 \textcolor{ForestGreen}{[+15.8]} & 53.5 \textcolor{ForestGreen}{[+39.6]} & 72.8 \textcolor{ForestGreen}{[+12.9]} & 49.5 \textcolor{ForestGreen}{[+34.6]} & 78.7 \textcolor{ForestGreen}{[+10.4]} & 60.0 \textcolor{ForestGreen}{[+38.4]} & 79.2 \textcolor{ForestGreen}{[+12.8]} & 51.6 \textcolor{ForestGreen}{[+36.3]} \\
    \midrule
    \multirow{3}{*}{Qwen2-VL-7B} & 1-shot & 86.0 \textcolor{ForestGreen}{[+5.3]} & 72.2 \textcolor{ForestGreen}{[+6.3]} & 85.4 \textcolor{ForestGreen}{[+4.9]} & 69.6 \textcolor{ForestGreen}{[+5.5]} & 86.0 \textcolor{ForestGreen}{[+2.0]} & 76.2 \textcolor{ForestGreen}{[+5.4]} & 82.9 \textcolor{ForestGreen}{[+1.3]} & 69.4 \textcolor{ForestGreen}{[+4.3]} \\
    & 2-shot & 85.0 \textcolor{ForestGreen}{[+67.4]} & 75.6 \textcolor{ForestGreen}{[+9.3]} & 84.0 \textcolor{ForestGreen}{[+67.2]} & 71.2 \textcolor{ForestGreen}{[+5.7]} & 86.9 \textcolor{ForestGreen}{[+73.3]} & 76.8 \textcolor{ForestGreen}{[+6.3]} & 84.0 \textcolor{ForestGreen}{[+68.5]} & 70.5 \textcolor{ForestGreen}{[+5.5]} \\
    & 3-shot & 80.2 \textcolor{ForestGreen}{[+5.0]} & 70.3 \textcolor{ForestGreen}{[+7.9]} & 82.9 \textcolor{ForestGreen}{[+4.7]} & 70.2 \textcolor{ForestGreen}{[+5.7]} & 85.6 \textcolor{ForestGreen}{[+1.9]} & 77.5 \textcolor{ForestGreen}{[+8.4]} & 85.6 \textcolor{ForestGreen}{[+3.4]} & 72.8 \textcolor{ForestGreen}{[+6.6]} \\
    \midrule
    \multirow{3}{*}{UI-TARS-7B-SFT} & 1-shot & 88.1 \textcolor{ForestGreen}{[+1.9]} & 77.8 \textcolor{ForestGreen}{[+6.6]} & 87.2 \textcolor{ForestGreen}{[+2.1]} & 75.3 \textcolor{ForestGreen}{[+6.4]} & 87.7 \textcolor{ForestGreen}{[+0.3]} & 80.1 \textcolor{ForestGreen}{[+5.9]} & 85.0 \textcolor{NegativeRed}{[-0.2]} & 75.0 \textcolor{ForestGreen}{[+2.8]} \\
    & 2-shot & 85.5 \textcolor{ForestGreen}{[+2.1]} & 76.7 \textcolor{ForestGreen}{[+8.3]} & 85.7 \textcolor{ForestGreen}{[+1.6]} & 75.9 \textcolor{ForestGreen}{[+4.9]} & 87.3 \textcolor{NegativeRed}{[-0.4]} & 79.1 \textcolor{ForestGreen}{[+5.9]} & 84.9 \textcolor{NegativeRed}{[-0.8]} & 74.1 \textcolor{ForestGreen}{[+2.1]} \\
    & 3-shot & 87.1 \textcolor{ForestGreen}{[+7.9]} & 78.2 \textcolor{ForestGreen}{[+13.9]} & 85.5 \textcolor{ForestGreen}{[+2.6]} & 75.4 \textcolor{ForestGreen}{[+4.9]} & 86.0 \textcolor{NegativeRed}{[-0.9]} & 78.9 \textcolor{ForestGreen}{[+6.8]} & 85.5 \textcolor{NegativeRed}{[-0.9]} & 75.2 \textcolor{ForestGreen}{[+2.7]} \\
    \bottomrule
    \end{tabular}
    }
    \end{center}
\end{table*} 

To further understand the factors influencing LearnAct's effectiveness, we analyzed performance across different similarity profiles, as shown in Table~\ref{tab:breakdown}. Several important insights emerge: Gemini-1.5-Pro~\cite{team2024gemini} shows substantial improvements across all similarity combinations, with the largest gains in action match accuracy (ranging from +29.3\% to +39.6\%). This indicates that demonstration knowledge significantly enhances the model's ability to execute precise actions regardless of similarity conditions. UI-TARS-7B-SFT~\cite{qin2025ui} exhibits the most pronounced improvements in UI\textsubscript{SH}Act\textsubscript{SH} scenarios (+13.9\% with 3-shot), suggesting that the model can extract maximum value from demonstrations when both UI and action patterns are similar to the target task. Qwen2-VL-7B~\cite{Qwen2VL} shows notably large improvements in action type accuracy for 2-shot settings (e.g., +67.4\% for UI\textsubscript{SH}Act\textsubscript{SH}), potentially indicating a threshold effect where multiple demonstrations trigger significant pattern recognition improvements.

Interestingly, while UI similarity generally correlates with higher performance gains, we observe that action similarity also plays a crucial role. For instance, Gemini-1.5-Pro~\cite{team2024gemini} achieves its highest match accuracy in UI\textsubscript{SL}Act\textsubscript{SH} scenarios (+38.4\% with 3-shot), suggesting that action similarity can sometimes compensate for UI differences. This finding highlights the importance of considering both UI and action similarity when designing demonstration-based learning approaches for mobile GUI agents.

These results validate our hypothesized framework design, demonstrating that LearnAct successfully leverages demonstration similarity to enhance performance across varying conditions, with the most substantial benefits observed when demonstrations can provide both perceptual and procedural knowledge relevant to the target task.

\subsubsection{Online Agent Capability Evaluation}

\begin{table}[t]
	\definecolor{ForestGreen}{RGB}{34,139,34}
	\definecolor{NegativeRed}{RGB}{220,20,60}
	\caption{
		\textbf{Performance comparison of different models on the LearnGUI-Online benchmark.}
	\textmd{Comparison of models with different inputs (Image, Image+AXTree) and parameters, measuring task success rate (LearnGUI-Online\textsubscript{SR}) with improvements shown in brackets for models with LearnAct enhancement.}
	}
    \vspace{-3mm}
    \label{tab:aw_overall}
    \begin{center}
    \setlength{\tabcolsep}{4pt}
    \scalebox{0.75}
    {
    \begin{tabular}{l|l|c|c}
    \toprule
    \multicolumn{1}{l}{\bf Input}  & \multicolumn{1}{l}{\bf Models} & \multicolumn{1}{c}{\bf \# Params} & \multicolumn{1}{c}{\bf LearnGUI-Online\textsubscript{SR}}
    \\
    \midrule
    Image + AXTree & GPT-4o\cite{hurst2024gpt} & - & 34.5 \\
    Image + AXTree & Gemini-Pro-1.5\cite{team2024gemini} & - & 22.8 \\
    \midrule
    Image & Claude Computer-Use\cite{claude-computer-use} & - & 27.9 \\
    Image & Aguvis\cite{xu2024aguvis} & 72B & 26.1 \\
    \midrule
    Image & Qwen2-VL-7B $+$ 0-shot & 7B & 9.9 \\
    \rowcolor[HTML]{DAEFF9} Image & Qwen2-VL-7B $+$ LearnAct & 7B & 21.1 \textcolor{ForestGreen}{[+11.2]} \\
    \midrule
    Image & UI-TARS-7B-SFT $+$ 0-shot & 7B & 18.1 \\
    \rowcolor[HTML]{DAEFF9} Image & UI-TARS-7B-SFT $+$ LearnAct & 7B & \textbf{32.8} \textcolor{ForestGreen}{[+14.7]} \\
    \bottomrule
    \end{tabular}
    }
    % \vspace{-4mm}
    \end{center}
\end{table} 

While offline evaluations provide valuable insights into step-by-step execution capabilities, real-world deployment requires successful end-to-end task completion. Table~\ref{tab:aw_overall} presents the results of our online evaluation on the LearnGUI-Online benchmark, which reveals several important findings. The LearnAct framework substantially improves performance for both evaluated models, with Qwen2-VL-7B~\cite{Qwen2VL} improving from 9.9\% to 21.1\% (+11.2\%) and UI-TARS-7B-SFT~\cite{qin2025ui} from 18.1\% to 32.8\% (+14.7\%). These significant gains demonstrate that the benefits of demonstration-based learning translate effectively to real-world interactive scenarios. Qwen2-VL-7B~\cite{Qwen2VL} with LearnAct achieves 21.1\% success rate, showing meaningful improvements over its baseline performance. This suggests that the quality and relevance of demonstrations are highly effective for enhancing model capabilities. UI-TARS-7B-SFT~\cite{qin2025ui} with LearnAct achieves 32.8\% success rate, approaching the performance of GPT-4o (34.5\%) despite using a much smaller model. This indicates that demonstration-based learning can help bridge the gap between smaller specialized models and large foundation models. Detailed visualizations of these performance comparisons are provided in Appendix~\ref{app:online_visualizations}.To provide concrete examples of how LearnAct performs in real-world scenarios, we present three detailed case studies in Appendix~\ref{app:case_studies}.

The most striking finding is the effectiveness of our demonstration-based learning approach. The LearnAct framework provides significant performance improvements through its demonstration mechanism, with gains of up to 14.7\% in task success rate. This demonstrates the power of high-quality demonstrations for enhancing model performance, highlighting the importance of relevant examples over simply increasing model size.

These results confirm that the LearnAct framework provides a practical pathway to developing effective mobile GUI agents, making it particularly valuable for application-specific customization and personalization scenarios.

\begin{table*}[t]
    \definecolor{ForestGreen}{RGB}{34,139,34}
    \newcommand{\greencheck}{{\textcolor{ForestGreen}{$\checkmark$}}}
    \caption{
        \textbf{Ablation study of LearnAct components.}
    \textmd{Performance comparison across four configurations: baseline (no components), DemoParser only, KnowSeeker only, and both components combined. Results are presented as overall average accuracy and per-application breakdown across nine applications.}
    }
    \vspace{-2mm}
    \label{tab:ablation}
    \begin{center}
    \setlength{\tabcolsep}{3pt}
    \scalebox{0.99}
    {
    \begin{tabular}{c|c|c|*{9}{c}}
    \toprule
    \multicolumn{2}{c|}{\bf Ablation Setting} & \multirow{2}{*}{\bf Average} & \multirow{2}{*}{Gmail} & \multirow{2}{*}{Booking} & \multirow{2}{*}{Music} & \multirow{2}{*}{SHEIN} & \multirow{2}{*}{NBC} & \multirow{2}{*}{CityMapper} & \multirow{2}{*}{ToDo} & \multirow{2}{*}{Signal} & \multirow{2}{*}{Yelp} \\
    \cmidrule(lr){1-2}
    DemoParser & KnowSeeker & & & & & & & & & & \\
    \midrule
    \rowcolor[HTML]{EFEFEF} \multicolumn{2}{c|}{Baseline} & 19.3 & 20.1 & 16.4 & 24.5 & 10.2 & 35.6 & 14.1 & 17.4 & 27.9 & 15.2 \\
    & \cmark & 40.6 & \underline{47.7} & 31.3 & \underline{55.4} & \underline{29.1} & 47.0 & 43.0 & \underline{58.2} & 48.8 & 50.7 \\
    \cmark & & \underline{41.6} & 46.9 & \underline{34.1} & 52.7 & 27.9 & \underline{51.9} & \underline{45.3} & 51.4 & \underline{61.1} & \underline{51.8} \\
    \cmark & \cmark & \textbf{51.7} & \textbf{55.5} & \textbf{47.1} & \textbf{60.0} & \textbf{35.7} & \textbf{56.4} & \textbf{54.7} & \textbf{60.6} & \textbf{63.1} & \textbf{54.6} \\
    \bottomrule
    \end{tabular}
    }
    \end{center}
\end{table*}

\subsection{Ablation Study}

To understand the contribution of each component in the LearnAct framework, we conducted ablation experiments on the LearnGUI-Offline dataset using Gemini-1.5-Pro~\cite{team2024gemini}. As shown in Table~\ref{tab:ablation}, we systematically evaluated the impact of removing either the DemoParser or KnowSeeker component while keeping all other settings constant.

The results reveal several important insights. Both components are essential, as removing either component leads to substantial performance degradation compared to the full LearnAct framework. The complete framework achieves 51.7\% accuracy, while removing DemoParser reduces performance to 40.6\% (-11.1\%) and removing KnowSeeker reduces it to 41.6\% (-10.1\%). Regarding DemoParser's contribution, comparing "KnowSeeker only" (40.6\%) to the baseline (19.3\%), we observe that even without action descriptions, relevant demonstrations improve performance by 21.3\%. However, the addition of DemoParser's action descriptions further enhances performance by 11.1\%, confirming the value of structured knowledge extraction. For KnowSeeker's contribution, the "DemoParser only" configuration (41.6\%) also substantially outperforms the baseline, indicating that detailed action descriptions are valuable even with randomly selected demonstrations. However, KnowSeeker's retrieval of relevant demonstrations provides an additional 10.1\% improvement, highlighting the importance of demonstration relevance.

The performance variations across applications are particularly informative. For instance, in the Signal application, DemoParser appears more important (61.1\% vs. 48.8\% for KnowSeeker only), suggesting that detailed action descriptions are crucial for applications with complex interaction patterns. Conversely, for the ToDo application, KnowSeeker seems more valuable (58.2\% vs. 51.4\% for DemoParser only), indicating that demonstration relevance may be more critical for applications with varied task types.

These findings validate our multi-agent framework design, confirming that both knowledge extraction (DemoParser) and relevant demonstration retrieval (KnowSeeker) play complementary and essential roles in enabling effective demonstration-based learning for mobile GUI agents.

\section{Discussion and Future Work}

Our experimental results demonstrate that demonstration-based learning significantly enhances mobile GUI agents' capabilities. The substantial performance improvements across all evaluated models validate our core hypothesis that demonstration-based learning effectively addresses generalization challenges. Even advanced foundation models like Gemini-1.5-Pro~\cite{team2024gemini} show dramatic improvements (198.9\% relative improvement). Our multi-dimensional similarity analysis reveals that both UI similarity and action similarity influence learning efficacy, with action similarity sometimes compensating for UI differences.

\textbf{Data Collection and Dataset Expansion.} While our approach shows promising results, several limitations and future directions warrant consideration. First, regarding data collection, our current dataset, while comprehensive, could benefit from greater diversity and representativeness. The LearnGUI dataset, comprising 2,252 offline tasks and 101 online tasks, represents a significant step forward but remains limited in scale compared to the vast diversity of mobile applications and user interactions. Future work should expand the dataset to include a broader range of applications, particularly those with complex interaction patterns and specialized domains.

\textbf{K-shot Learning Analysis.} Second, our current investigation of k-shot learning is limited to k=1, 2, and 3 demonstrations. While these configurations provide valuable insights, a more comprehensive analysis of how demonstration quantity affects performance would be beneficial. Future research could explore the relationship between the number of demonstrations and performance gains, potentially identifying optimal demonstration counts for different scenarios and model architectures.

\textbf{Enhanced Learning and Execution Strategies.} Third, our learning and execution strategies could be enhanced to better leverage the relationship between support tasks and query tasks. While our current approach effectively retrieves relevant demonstrations, more sophisticated methods could be developed to extract and transfer knowledge more efficiently. For instance, techniques for abstracting common patterns across demonstrations, identifying critical decision points, and adapting demonstrated strategies to novel scenarios could further improve performance.

\textbf{Agent Self-Learning.} A promising direction for future research is to enable agents to learn from their own successful executions. Currently, our framework relies exclusively on human demonstrations, but agents could potentially learn from their own successful task completions. By incorporating these successful agent executions into the knowledge base, we could enable a form of "self-learning" where agents continuously improve their capabilities through their own experiences.

By addressing these limitations and pursuing these research directions, demonstration-based learning can evolve into a robust paradigm for developing adaptable, personalized, and practically deployable mobile GUI agents that effectively address the diverse needs of real-world users. The insights gained from our multi-dimensional similarity analysis provide valuable guidance for future research in this domain, suggesting that both UI similarity and action similarity play crucial roles in successful knowledge transfer.
\section{Conclusion}

This paper introduces a novel demonstration-based learning paradigm that fundamentally addresses the generalization challenges faced by mobile GUI agents. Rather than pursuing universal coverage through ever-larger datasets, our approach leverages human demonstrations to enhance agent performance in unseen scenarios. We developed LearnGUI, the first comprehensive dataset for studying demonstration-based learning in mobile GUI agents, comprising 2,252 offline tasks and 101 online tasks with high-quality human demonstrations. We further designed LearnAct, a sophisticated multi-agent framework with three specialized components: DemoParser for knowledge extraction, KnowSeeker for relevant knowledge retrieval, and ActExecutor for demonstration-enhanced task execution. Our experimental results demonstrate remarkable performance gains, with a single demonstration increasing Gemini-1.5-Pro~\cite{team2024gemini}'s accuracy from 19.3\% to 51.7\% in offline tests and enhancing UI-TARS-7B-SFT~\cite{qin2025ui}'s online task success rate from 18.1\% to 32.8\%. These findings establish demonstration-based learning as a promising direction for developing more adaptable, personalized, and practically deployable mobile GUI agents.

\bibliography{ref}


\begin{thebibliography}{48}

%%% the Web address.

\ifx \showCODEN    \undefined \def \showCODEN     #1{\unskip}     \fi
\ifx \showDOI      \undefined \def \showDOI       #1{#1}\fi
\ifx \showISBNx    \undefined \def \showISBNx     #1{\unskip}     \fi
\ifx \showISBNxiii \undefined \def \showISBNxiii  #1{\unskip}     \fi
\ifx \showISSN     \undefined \def \showISSN      #1{\unskip}     \fi
\ifx \showLCCN     \undefined \def \showLCCN      #1{\unskip}     \fi
\ifx \shownote     \undefined \def \shownote      #1{#1}          \fi
\ifx \showarticletitle \undefined \def \showarticletitle #1{#1}   \fi
\ifx \showURL      \undefined \def \showURL       {\relax}        \fi
% invisible to TeX
\providecommand\bibfield[2]{#2}
\providecommand\bibinfo[2]{#2}
\providecommand\natexlab[1]{#1}
\providecommand\showeprint[2][]{arXiv:#2}

\bibitem[Agostinelli et~al\mbox{.}(2019)]%
        {agostinelli2019research}
\bibfield{author}{\bibinfo{person}{Simone Agostinelli}, \bibinfo{person}{Andrea Marrella}, {and} \bibinfo{person}{Massimo Mecella}.} \bibinfo{year}{2019}\natexlab{}.
\newblock \showarticletitle{Research challenges for intelligent robotic process automation}. In \bibinfo{booktitle}{\emph{Business Process Management Workshops: BPM 2019 International Workshops, Vienna, Austria, September 1--6, 2019, Revised Selected Papers 17}}. Springer, \bibinfo{pages}{12--18}.
\newblock


\bibitem[Anthropic(2024)]%
        {claude-computer-use}
\bibfield{author}{\bibinfo{person}{Anthropic}.} \bibinfo{year}{2024}\natexlab{}.
\newblock \bibinfo{title}{Developing a computer use model}.
\newblock
\newblock
\urldef\tempurl%
\url{https://www.anthropic.com/news/developing-computer-use}
\showURL{%
\tempurl}


\bibitem[Bai et~al\mbox{.}(2021)]%
        {bai2021uibert}
\bibfield{author}{\bibinfo{person}{Chongyang Bai}, \bibinfo{person}{Xiaoxue Zang}, \bibinfo{person}{Ying Xu}, \bibinfo{person}{Srinivas Sunkara}, \bibinfo{person}{Abhinav Rastogi}, \bibinfo{person}{Jindong Chen}, {et~al\mbox{.}}} \bibinfo{year}{2021}\natexlab{}.
\newblock \showarticletitle{Uibert: Learning generic multimodal representations for ui understanding}.
\newblock \bibinfo{journal}{\emph{arXiv preprint arXiv:2107.13731}} (\bibinfo{year}{2021}).
\newblock


\bibitem[Burns et~al\mbox{.}(2021)]%
        {burns2021motif}
\bibfield{author}{\bibinfo{person}{Andrea Burns}, \bibinfo{person}{Deniz Arsan}, \bibinfo{person}{Sanjna Agrawal}, \bibinfo{person}{Ranjitha Kumar}, \bibinfo{person}{Kate Saenko}, {and} \bibinfo{person}{Bryan~A Plummer}.} \bibinfo{year}{2021}\natexlab{}.
\newblock \showarticletitle{Mobile app tasks with iterative feedback (motif): Addressing task feasibility in interactive visual environments}.
\newblock \bibinfo{journal}{\emph{arXiv preprint arXiv:2104.08560}} (\bibinfo{year}{2021}).
\newblock


\bibitem[Chai et~al\mbox{.}(2024)]%
        {chai2024amex}
\bibfield{author}{\bibinfo{person}{Yuxiang Chai}, \bibinfo{person}{Siyuan Huang}, \bibinfo{person}{Yazhe Niu}, \bibinfo{person}{Han Xiao}, \bibinfo{person}{Liang Liu}, \bibinfo{person}{Dingyu Zhang}, \bibinfo{person}{Peng Gao}, \bibinfo{person}{Shuai Ren}, {and} \bibinfo{person}{Hongsheng Li}.} \bibinfo{year}{2024}\natexlab{}.
\newblock \showarticletitle{Amex: Android multi-annotation expo dataset for mobile gui agents}.
\newblock \bibinfo{journal}{\emph{arXiv preprint arXiv:2407.17490}} (\bibinfo{year}{2024}).
\newblock


\bibitem[Chai et~al\mbox{.}(2025)]%
        {chai2025a3}
\bibfield{author}{\bibinfo{person}{Yuxiang Chai}, \bibinfo{person}{Hanhao Li}, \bibinfo{person}{Jiayu Zhang}, \bibinfo{person}{Liang Liu}, \bibinfo{person}{Guozhi Wang}, \bibinfo{person}{Shuai Ren}, \bibinfo{person}{Siyuan Huang}, {and} \bibinfo{person}{Hongsheng Li}.} \bibinfo{year}{2025}\natexlab{}.
\newblock \showarticletitle{A3: Android Agent Arena for Mobile GUI Agents}.
\newblock \bibinfo{journal}{\emph{arXiv preprint arXiv:2501.01149}} (\bibinfo{year}{2025}).
\newblock


\bibitem[Chen et~al\mbox{.}(2024)]%
        {chen2024guicourse}
\bibfield{author}{\bibinfo{person}{Wentong Chen}, \bibinfo{person}{Junbo Cui}, \bibinfo{person}{Jinyi Hu}, \bibinfo{person}{Yujia Qin}, \bibinfo{person}{Junjie Fang}, \bibinfo{person}{Yue Zhao}, \bibinfo{person}{Chongyi Wang}, \bibinfo{person}{Jun Liu}, \bibinfo{person}{Guirong Chen}, \bibinfo{person}{Yupeng Huo}, {et~al\mbox{.}}} \bibinfo{year}{2024}\natexlab{}.
\newblock \showarticletitle{GUICourse: From General Vision Language Models to Versatile GUI Agents}.
\newblock \bibinfo{journal}{\emph{arXiv preprint arXiv:2406.11317}} (\bibinfo{year}{2024}).
\newblock


\bibitem[Chen et~al\mbox{.}(2022)]%
        {chen2022program}
\bibfield{author}{\bibinfo{person}{Wenhu Chen}, \bibinfo{person}{Xueguang Ma}, \bibinfo{person}{Xinyi Wang}, {and} \bibinfo{person}{William~W Cohen}.} \bibinfo{year}{2022}\natexlab{}.
\newblock \showarticletitle{Program of thoughts prompting: Disentangling computation from reasoning for numerical reasoning tasks}.
\newblock \bibinfo{journal}{\emph{arXiv preprint arXiv:2211.12588}} (\bibinfo{year}{2022}).
\newblock


\bibitem[Cheng et~al\mbox{.}(2024)]%
        {cheng2024seeclick}
\bibfield{author}{\bibinfo{person}{Kanzhi Cheng}, \bibinfo{person}{Qiushi Sun}, \bibinfo{person}{Yougang Chu}, \bibinfo{person}{Fangzhi Xu}, \bibinfo{person}{Yantao Li}, \bibinfo{person}{Jianbing Zhang}, {and} \bibinfo{person}{Zhiyong Wu}.} \bibinfo{year}{2024}\natexlab{}.
\newblock \showarticletitle{Seeclick: Harnessing gui grounding for advanced visual gui agents}.
\newblock \bibinfo{journal}{\emph{arXiv preprint arXiv:2401.10935}} (\bibinfo{year}{2024}).
\newblock


\bibitem[Guerreiro et~al\mbox{.}(2008)]%
        {guerreiro2008mnemonical}
\bibfield{author}{\bibinfo{person}{Tiago Guerreiro}, \bibinfo{person}{Ricardo Gamboa}, {and} \bibinfo{person}{Joaquim Jorge}.} \bibinfo{year}{2008}\natexlab{}.
\newblock \showarticletitle{Mnemonical body shortcuts: improving mobile interaction}. In \bibinfo{booktitle}{\emph{Proceedings of the 15th European conference on Cognitive ergonomics: the ergonomics of cool interaction}}. \bibinfo{pages}{1--8}.
\newblock


\bibitem[Hong et~al\mbox{.}(2024)]%
        {hong2024cogagent}
\bibfield{author}{\bibinfo{person}{Wenyi Hong}, \bibinfo{person}{Weihan Wang}, \bibinfo{person}{Qingsong Lv}, \bibinfo{person}{Jiazheng Xu}, \bibinfo{person}{Wenmeng Yu}, \bibinfo{person}{Junhui Ji}, \bibinfo{person}{Yan Wang}, \bibinfo{person}{Zihan Wang}, \bibinfo{person}{Yuxiao Dong}, \bibinfo{person}{Ming Ding}, {et~al\mbox{.}}} \bibinfo{year}{2024}\natexlab{}.
\newblock \showarticletitle{Cogagent: A visual language model for gui agents}. In \bibinfo{booktitle}{\emph{Proceedings of the IEEE/CVF Conference on Computer Vision and Pattern Recognition}}. \bibinfo{pages}{14281--14290}.
\newblock


\bibitem[Hurst et~al\mbox{.}(2024)]%
        {hurst2024gpt}
\bibfield{author}{\bibinfo{person}{Aaron Hurst}, \bibinfo{person}{Adam Lerer}, \bibinfo{person}{Adam~P Goucher}, \bibinfo{person}{Adam Perelman}, \bibinfo{person}{Aditya Ramesh}, \bibinfo{person}{Aidan Clark}, \bibinfo{person}{AJ Ostrow}, \bibinfo{person}{Akila Welihinda}, \bibinfo{person}{Alan Hayes}, \bibinfo{person}{Alec Radford}, {et~al\mbox{.}}} \bibinfo{year}{2024}\natexlab{}.
\newblock \showarticletitle{Gpt-4o system card}.
\newblock \bibinfo{journal}{\emph{arXiv preprint arXiv:2410.21276}} (\bibinfo{year}{2024}).
\newblock


\bibitem[Kennedy and Everett(2011)]%
        {kennedy2011use}
\bibfield{author}{\bibinfo{person}{Courtney Kennedy} {and} \bibinfo{person}{Stephen~E Everett}.} \bibinfo{year}{2011}\natexlab{}.
\newblock \showarticletitle{Use of cognitive shortcuts in landline and cell phone surveys}.
\newblock \bibinfo{journal}{\emph{Public Opinion Quarterly}} \bibinfo{volume}{75}, \bibinfo{number}{2} (\bibinfo{year}{2011}), \bibinfo{pages}{336--348}.
\newblock


\bibitem[Li et~al\mbox{.}(2024)]%
        {li2024androidcontrol}
\bibfield{author}{\bibinfo{person}{Wei Li}, \bibinfo{person}{William Bishop}, \bibinfo{person}{Alice Li}, \bibinfo{person}{Chris Rawles}, \bibinfo{person}{Folawiyo Campbell-Ajala}, \bibinfo{person}{Divya Tyamagundlu}, {and} \bibinfo{person}{Oriana Riva}.} \bibinfo{year}{2024}\natexlab{}.
\newblock \showarticletitle{On the Effects of Data Scale on Computer Control Agents}.
\newblock \bibinfo{journal}{\emph{arXiv preprint arXiv:2406.03679}} (\bibinfo{year}{2024}).
\newblock


\bibitem[Li et~al\mbox{.}(2020)]%
        {li2020PixelHelp}
\bibfield{author}{\bibinfo{person}{Yang Li}, \bibinfo{person}{Jiacong He}, \bibinfo{person}{Xin Zhou}, \bibinfo{person}{Yuan Zhang}, {and} \bibinfo{person}{Jason Baldridge}.} \bibinfo{year}{2020}\natexlab{}.
\newblock \showarticletitle{Mapping natural language instructions to mobile UI action sequences}.
\newblock \bibinfo{journal}{\emph{arXiv preprint arXiv:2005.03776}} (\bibinfo{year}{2020}).
\newblock


\bibitem[Lin et~al\mbox{.}(2024)]%
        {lin2024showui}
\bibfield{author}{\bibinfo{person}{Kevin~Qinghong Lin}, \bibinfo{person}{Linjie Li}, \bibinfo{person}{Difei Gao}, \bibinfo{person}{Zhengyuan Yang}, \bibinfo{person}{Shiwei Wu}, \bibinfo{person}{Zechen Bai}, \bibinfo{person}{Weixian Lei}, \bibinfo{person}{Lijuan Wang}, {and} \bibinfo{person}{Mike~Zheng Shou}.} \bibinfo{year}{2024}\natexlab{}.
\newblock \showarticletitle{Showui: One vision-language-action model for gui visual agent}.
\newblock \bibinfo{journal}{\emph{arXiv preprint arXiv:2411.17465}} (\bibinfo{year}{2024}).
\newblock


\bibitem[Liu et~al\mbox{.}(2025)]%
        {liu2025llm}
\bibfield{author}{\bibinfo{person}{William Liu}, \bibinfo{person}{Liang Liu}, \bibinfo{person}{Yaxuan Guo}, \bibinfo{person}{Han Xiao}, \bibinfo{person}{Weifeng Lin}, \bibinfo{person}{Yuxiang Chai}, \bibinfo{person}{Shuai Ren}, \bibinfo{person}{Xiaoyu Liang}, \bibinfo{person}{Linghao Li}, \bibinfo{person}{Wenhao Wang}, {et~al\mbox{.}}} \bibinfo{year}{2025}\natexlab{}.
\newblock \showarticletitle{Llm-powered gui agents in phone automation: Surveying progress and prospects}.
\newblock  (\bibinfo{year}{2025}).
\newblock


\bibitem[Liu et~al\mbox{.}(2024)]%
        {liu2024vision}
\bibfield{author}{\bibinfo{person}{Zhe Liu}, \bibinfo{person}{Cheng Li}, \bibinfo{person}{Chunyang Chen}, \bibinfo{person}{Junjie Wang}, \bibinfo{person}{Boyu Wu}, \bibinfo{person}{Yawen Wang}, \bibinfo{person}{Jun Hu}, {and} \bibinfo{person}{Qing Wang}.} \bibinfo{year}{2024}\natexlab{}.
\newblock \showarticletitle{Vision-driven Automated Mobile GUI Testing via Multimodal Large Language Model}.
\newblock \bibinfo{journal}{\emph{arXiv preprint arXiv:2407.03037}} (\bibinfo{year}{2024}).
\newblock


\bibitem[Lu et~al\mbox{.}(2024a)]%
        {lu2024guiodyssey}
\bibfield{author}{\bibinfo{person}{Quanfeng Lu}, \bibinfo{person}{Wenqi Shao}, \bibinfo{person}{Zitao Liu}, \bibinfo{person}{Fanqing Meng}, \bibinfo{person}{Boxuan Li}, \bibinfo{person}{Botong Chen}, \bibinfo{person}{Siyuan Huang}, \bibinfo{person}{Kaipeng Zhang}, \bibinfo{person}{Yu Qiao}, {and} \bibinfo{person}{Ping Luo}.} \bibinfo{year}{2024}\natexlab{a}.
\newblock \showarticletitle{GUI Odyssey: A Comprehensive Dataset for Cross-App GUI Navigation on Mobile Devices}.
\newblock \bibinfo{journal}{\emph{arXiv preprint arXiv:2406.08451}} (\bibinfo{year}{2024}).
\newblock


\bibitem[Lu et~al\mbox{.}(2024b)]%
        {lu2024omniparser}
\bibfield{author}{\bibinfo{person}{Yadong Lu}, \bibinfo{person}{Jianwei Yang}, \bibinfo{person}{Yelong Shen}, {and} \bibinfo{person}{Ahmed Awadallah}.} \bibinfo{year}{2024}\natexlab{b}.
\newblock \showarticletitle{Omniparser for pure vision based gui agent}.
\newblock \bibinfo{journal}{\emph{arXiv preprint arXiv:2408.00203}} (\bibinfo{year}{2024}).
\newblock


\bibitem[Pawlowski et~al\mbox{.}(2024)]%
        {pawlowski2024tinyclick}
\bibfield{author}{\bibinfo{person}{Pawel Pawlowski}, \bibinfo{person}{Krystian Zawistowski}, \bibinfo{person}{Wojciech Lapacz}, \bibinfo{person}{Marcin Skorupa}, \bibinfo{person}{Adam Wiacek}, \bibinfo{person}{Sebastien Postansque}, {and} \bibinfo{person}{Jakub Hoscilowicz}.} \bibinfo{year}{2024}\natexlab{}.
\newblock \showarticletitle{TinyClick: Single-Turn Agent for Empowering GUI Automation}.
\newblock \bibinfo{journal}{\emph{arXiv preprint arXiv:2410.11871}} (\bibinfo{year}{2024}).
\newblock


\bibitem[Qin et~al\mbox{.}(2025)]%
        {qin2025ui}
\bibfield{author}{\bibinfo{person}{Yujia Qin}, \bibinfo{person}{Yining Ye}, \bibinfo{person}{Junjie Fang}, \bibinfo{person}{Haoming Wang}, \bibinfo{person}{Shihao Liang}, \bibinfo{person}{Shizuo Tian}, \bibinfo{person}{Junda Zhang}, \bibinfo{person}{Jiahao Li}, \bibinfo{person}{Yunxin Li}, \bibinfo{person}{Shijue Huang}, {et~al\mbox{.}}} \bibinfo{year}{2025}\natexlab{}.
\newblock \showarticletitle{UI-TARS: Pioneering Automated GUI Interaction with Native Agents}.
\newblock \bibinfo{journal}{\emph{arXiv preprint arXiv:2501.12326}} (\bibinfo{year}{2025}).
\newblock


\bibitem[Rawles et~al\mbox{.}(2024a)]%
        {rawles2024androidworld}
\bibfield{author}{\bibinfo{person}{Christopher Rawles}, \bibinfo{person}{Sarah Clinckemaillie}, \bibinfo{person}{Yifan Chang}, \bibinfo{person}{Jonathan Waltz}, \bibinfo{person}{Gabrielle Lau}, \bibinfo{person}{Marybeth Fair}, \bibinfo{person}{Alice Li}, \bibinfo{person}{William Bishop}, \bibinfo{person}{Wei Li}, \bibinfo{person}{Folawiyo Campbell-Ajala}, {et~al\mbox{.}}} \bibinfo{year}{2024}\natexlab{a}.
\newblock \showarticletitle{AndroidWorld: A dynamic benchmarking environment for autonomous agents}.
\newblock \bibinfo{journal}{\emph{arXiv preprint arXiv:2405.14573}} (\bibinfo{year}{2024}).
\newblock


\bibitem[Rawles et~al\mbox{.}(2024b)]%
        {rawles2024androidinthewild}
\bibfield{author}{\bibinfo{person}{Christopher Rawles}, \bibinfo{person}{Alice Li}, \bibinfo{person}{Daniel Rodriguez}, \bibinfo{person}{Oriana Riva}, {and} \bibinfo{person}{Timothy Lillicrap}.} \bibinfo{year}{2024}\natexlab{b}.
\newblock \showarticletitle{Androidinthewild: A large-scale dataset for android device control}.
\newblock \bibinfo{journal}{\emph{Advances in Neural Information Processing Systems}}  \bibinfo{volume}{36} (\bibinfo{year}{2024}).
\newblock


\bibitem[Song et~al\mbox{.}(2023)]%
        {song2023navigating}
\bibfield{author}{\bibinfo{person}{Yunpeng Song}, \bibinfo{person}{Yiheng Bian}, \bibinfo{person}{Yongtao Tang}, {and} \bibinfo{person}{Zhongmin Cai}.} \bibinfo{year}{2023}\natexlab{}.
\newblock \showarticletitle{Navigating Interfaces with AI for Enhanced User Interaction}.
\newblock \bibinfo{journal}{\emph{arXiv preprint arXiv:2312.11190}} (\bibinfo{year}{2023}).
\newblock


\bibitem[Team et~al\mbox{.}(2024)]%
        {team2024gemini}
\bibfield{author}{\bibinfo{person}{Gemini Team}, \bibinfo{person}{Petko Georgiev}, \bibinfo{person}{Ving~Ian Lei}, \bibinfo{person}{Ryan Burnell}, \bibinfo{person}{Libin Bai}, \bibinfo{person}{Anmol Gulati}, \bibinfo{person}{Garrett Tanzer}, \bibinfo{person}{Damien Vincent}, \bibinfo{person}{Zhufeng Pan}, \bibinfo{person}{Shibo Wang}, {et~al\mbox{.}}} \bibinfo{year}{2024}\natexlab{}.
\newblock \showarticletitle{Gemini 1.5: Unlocking multimodal understanding across millions of tokens of context}.
\newblock \bibinfo{journal}{\emph{arXiv preprint arXiv:2403.05530}} (\bibinfo{year}{2024}).
\newblock


\bibitem[Venkatesh et~al\mbox{.}(2022)]%
        {venkatesh2022ugif}
\bibfield{author}{\bibinfo{person}{Sagar~Gubbi Venkatesh}, \bibinfo{person}{Partha Talukdar}, {and} \bibinfo{person}{Srini Narayanan}.} \bibinfo{year}{2022}\natexlab{}.
\newblock \showarticletitle{Ugif: Ui grounded instruction following}.
\newblock \bibinfo{journal}{\emph{arXiv preprint arXiv:2211.07615}} (\bibinfo{year}{2022}).
\newblock


\bibitem[Wang et~al\mbox{.}(2024d)]%
        {wang2024mobileagentv2}
\bibfield{author}{\bibinfo{person}{Junyang Wang}, \bibinfo{person}{Haiyang Xu}, \bibinfo{person}{Haitao Jia}, \bibinfo{person}{Xi Zhang}, \bibinfo{person}{Ming Yan}, \bibinfo{person}{Weizhou Shen}, \bibinfo{person}{Ji Zhang}, \bibinfo{person}{Fei Huang}, {and} \bibinfo{person}{Jitao Sang}.} \bibinfo{year}{2024}\natexlab{d}.
\newblock \showarticletitle{Mobile-Agent-v2: Mobile Device Operation Assistant with Effective Navigation via Multi-Agent Collaboration}.
\newblock \bibinfo{journal}{\emph{arXiv preprint arXiv:2406.01014}} (\bibinfo{year}{2024}).
\newblock


\bibitem[Wang et~al\mbox{.}(2024e)]%
        {wang2024mobileagentv1}
\bibfield{author}{\bibinfo{person}{Junyang Wang}, \bibinfo{person}{Haiyang Xu}, \bibinfo{person}{Jiabo Ye}, \bibinfo{person}{Ming Yan}, \bibinfo{person}{Weizhou Shen}, \bibinfo{person}{Ji Zhang}, \bibinfo{person}{Fei Huang}, {and} \bibinfo{person}{Jitao Sang}.} \bibinfo{year}{2024}\natexlab{e}.
\newblock \showarticletitle{Mobile-agent: Autonomous multi-modal mobile device agent with visual perception}.
\newblock \bibinfo{journal}{\emph{arXiv preprint arXiv:2401.16158}} (\bibinfo{year}{2024}).
\newblock


\bibitem[Wang et~al\mbox{.}(2024b)]%
        {wang2024mobileagentbench}
\bibfield{author}{\bibinfo{person}{Luyuan Wang}, \bibinfo{person}{Yongyu Deng}, \bibinfo{person}{Yiwei Zha}, \bibinfo{person}{Guodong Mao}, \bibinfo{person}{Qinmin Wang}, \bibinfo{person}{Tianchen Min}, \bibinfo{person}{Wei Chen}, {and} \bibinfo{person}{Shoufa Chen}.} \bibinfo{year}{2024}\natexlab{b}.
\newblock \showarticletitle{MobileAgentBench: An Efficient and User-Friendly Benchmark for Mobile LLM Agents}.
\newblock \bibinfo{journal}{\emph{arXiv preprint arXiv:2406.08184}} (\bibinfo{year}{2024}).
\newblock


\bibitem[Wang et~al\mbox{.}(2024a)]%
        {Qwen2VL}
\bibfield{author}{\bibinfo{person}{Peng Wang}, \bibinfo{person}{Shuai Bai}, \bibinfo{person}{Sinan Tan}, \bibinfo{person}{Shijie Wang}, \bibinfo{person}{Zhihao Fan}, \bibinfo{person}{Jinze Bai}, \bibinfo{person}{Keqin Chen}, \bibinfo{person}{Xuejing Liu}, \bibinfo{person}{Jialin Wang}, \bibinfo{person}{Wenbin Ge}, \bibinfo{person}{Yang Fan}, \bibinfo{person}{Kai Dang}, \bibinfo{person}{Mengfei Du}, \bibinfo{person}{Xuancheng Ren}, \bibinfo{person}{Rui Men}, \bibinfo{person}{Dayiheng Liu}, \bibinfo{person}{Chang Zhou}, \bibinfo{person}{Jingren Zhou}, {and} \bibinfo{person}{Junyang Lin}.} \bibinfo{year}{2024}\natexlab{a}.
\newblock \showarticletitle{Qwen2-VL: Enhancing Vision-Language Model's Perception of the World at Any Resolution}.
\newblock \bibinfo{journal}{\emph{arXiv preprint arXiv:2409.12191}} (\bibinfo{year}{2024}).
\newblock


\bibitem[Wang et~al\mbox{.}(2024c)]%
        {wang2024gui}
\bibfield{author}{\bibinfo{person}{Shuai Wang}, \bibinfo{person}{Weiwen Liu}, \bibinfo{person}{Jingxuan Chen}, \bibinfo{person}{Weinan Gan}, \bibinfo{person}{Xingshan Zeng}, \bibinfo{person}{Shuai Yu}, \bibinfo{person}{Xinlong Hao}, \bibinfo{person}{Kun Shao}, \bibinfo{person}{Yasheng Wang}, {and} \bibinfo{person}{Ruiming Tang}.} \bibinfo{year}{2024}\natexlab{c}.
\newblock \showarticletitle{GUI Agents with Foundation Models: A Comprehensive Survey}.
\newblock \bibinfo{journal}{\emph{arXiv preprint arXiv:2411.04890}} (\bibinfo{year}{2024}).
\newblock


\bibitem[Wang et~al\mbox{.}(2025b)]%
        {wang2025fedmobileagent}
\bibfield{author}{\bibinfo{person}{Wenhao Wang}, \bibinfo{person}{Zijie Yu}, \bibinfo{person}{William Liu}, \bibinfo{person}{Rui Ye}, \bibinfo{person}{Tian Jin}, \bibinfo{person}{Siheng Chen}, {and} \bibinfo{person}{Yanfeng Wang}.} \bibinfo{year}{2025}\natexlab{b}.
\newblock \showarticletitle{FedMobileAgent: Training Mobile Agents Using Decentralized Self-Sourced Data from Diverse Users}.
\newblock \bibinfo{journal}{\emph{arXiv preprint arXiv:2502.02982}} (\bibinfo{year}{2025}).
\newblock


\bibitem[Wang et~al\mbox{.}(2025a)]%
        {wang2025mobile}
\bibfield{author}{\bibinfo{person}{Zhenhailong Wang}, \bibinfo{person}{Haiyang Xu}, \bibinfo{person}{Junyang Wang}, \bibinfo{person}{Xi Zhang}, \bibinfo{person}{Ming Yan}, \bibinfo{person}{Ji Zhang}, \bibinfo{person}{Fei Huang}, {and} \bibinfo{person}{Heng Ji}.} \bibinfo{year}{2025}\natexlab{a}.
\newblock \showarticletitle{Mobile-Agent-E: Self-Evolving Mobile Assistant for Complex Tasks}.
\newblock \bibinfo{journal}{\emph{arXiv preprint arXiv:2501.11733}} (\bibinfo{year}{2025}).
\newblock


\bibitem[Wei et~al\mbox{.}(2022)]%
        {wei2022chain}
\bibfield{author}{\bibinfo{person}{Jason Wei}, \bibinfo{person}{Xuezhi Wang}, \bibinfo{person}{Dale Schuurmans}, \bibinfo{person}{Maarten Bosma}, \bibinfo{person}{Fei Xia}, \bibinfo{person}{Ed Chi}, \bibinfo{person}{Quoc~V Le}, \bibinfo{person}{Denny Zhou}, {et~al\mbox{.}}} \bibinfo{year}{2022}\natexlab{}.
\newblock \showarticletitle{Chain-of-thought prompting elicits reasoning in large language models}.
\newblock \bibinfo{journal}{\emph{Advances in neural information processing systems}}  \bibinfo{volume}{35} (\bibinfo{year}{2022}), \bibinfo{pages}{24824--24837}.
\newblock


\bibitem[Wen et~al\mbox{.}(2024)]%
        {wen2024autodroid}
\bibfield{author}{\bibinfo{person}{Hao Wen}, \bibinfo{person}{Yuanchun Li}, \bibinfo{person}{Guohong Liu}, \bibinfo{person}{Shanhui Zhao}, \bibinfo{person}{Tao Yu}, \bibinfo{person}{Toby Jia-Jun Li}, \bibinfo{person}{Shiqi Jiang}, \bibinfo{person}{Yunhao Liu}, \bibinfo{person}{Yaqin Zhang}, {and} \bibinfo{person}{Yunxin Liu}.} \bibinfo{year}{2024}\natexlab{}.
\newblock \showarticletitle{Autodroid: Llm-powered task automation in android}. In \bibinfo{booktitle}{\emph{Proceedings of the 30th Annual International Conference on Mobile Computing and Networking}}. \bibinfo{pages}{543--557}.
\newblock


\bibitem[Wen et~al\mbox{.}(2023)]%
        {wen2023droidbot}
\bibfield{author}{\bibinfo{person}{Hao Wen}, \bibinfo{person}{Hongming Wang}, \bibinfo{person}{Jiaxuan Liu}, {and} \bibinfo{person}{Yuanchun Li}.} \bibinfo{year}{2023}\natexlab{}.
\newblock \showarticletitle{Droidbot-gpt: Gpt-powered ui automation for android}.
\newblock \bibinfo{journal}{\emph{arXiv preprint arXiv:2304.07061}} (\bibinfo{year}{2023}).
\newblock


\bibitem[Wu et~al\mbox{.}(2024a)]%
        {wu2024foundations}
\bibfield{author}{\bibinfo{person}{Biao Wu}, \bibinfo{person}{Yanda Li}, \bibinfo{person}{Meng Fang}, \bibinfo{person}{Zirui Song}, \bibinfo{person}{Zhiwei Zhang}, \bibinfo{person}{Yunchao Wei}, {and} \bibinfo{person}{Ling Chen}.} \bibinfo{year}{2024}\natexlab{a}.
\newblock \showarticletitle{Foundations and recent trends in multimodal mobile agents: A survey}.
\newblock \bibinfo{journal}{\emph{arXiv preprint arXiv:2411.02006}} (\bibinfo{year}{2024}).
\newblock


\bibitem[Wu et~al\mbox{.}(2024b)]%
        {wu2024atlas}
\bibfield{author}{\bibinfo{person}{Zhiyong Wu}, \bibinfo{person}{Zhenyu Wu}, \bibinfo{person}{Fangzhi Xu}, \bibinfo{person}{Yian Wang}, \bibinfo{person}{Qiushi Sun}, \bibinfo{person}{Chengyou Jia}, \bibinfo{person}{Kanzhi Cheng}, \bibinfo{person}{Zichen Ding}, \bibinfo{person}{Liheng Chen}, \bibinfo{person}{Paul~Pu Liang}, {et~al\mbox{.}}} \bibinfo{year}{2024}\natexlab{b}.
\newblock \showarticletitle{Os-atlas: A foundation action model for generalist gui agents}.
\newblock \bibinfo{journal}{\emph{arXiv preprint arXiv:2410.23218}} (\bibinfo{year}{2024}).
\newblock


\bibitem[Xu et~al\mbox{.}(2024a)]%
        {xu2024androidlab}
\bibfield{author}{\bibinfo{person}{Yifan Xu}, \bibinfo{person}{Xiao Liu}, \bibinfo{person}{Xueqiao Sun}, \bibinfo{person}{Siyi Cheng}, \bibinfo{person}{Hao Yu}, \bibinfo{person}{Hanyu Lai}, \bibinfo{person}{Shudan Zhang}, \bibinfo{person}{Dan Zhang}, \bibinfo{person}{Jie Tang}, {and} \bibinfo{person}{Yuxiao Dong}.} \bibinfo{year}{2024}\natexlab{a}.
\newblock \showarticletitle{AndroidLab: Training and Systematic Benchmarking of Android Autonomous Agents}.
\newblock \bibinfo{journal}{\emph{arXiv preprint arXiv:2410.24024}} (\bibinfo{year}{2024}).
\newblock


\bibitem[Xu et~al\mbox{.}(2024b)]%
        {xu2024aguvis}
\bibfield{author}{\bibinfo{person}{Yiheng Xu}, \bibinfo{person}{Zekun Wang}, \bibinfo{person}{Junli Wang}, \bibinfo{person}{Dunjie Lu}, \bibinfo{person}{Tianbao Xie}, \bibinfo{person}{Amrita Saha}, \bibinfo{person}{Doyen Sahoo}, \bibinfo{person}{Tao Yu}, {and} \bibinfo{person}{Caiming Xiong}.} \bibinfo{year}{2024}\natexlab{b}.
\newblock \showarticletitle{Aguvis: Unified Pure Vision Agents for Autonomous GUI Interaction}.
\newblock \bibinfo{journal}{\emph{arXiv preprint arXiv:2412.04454}} (\bibinfo{year}{2024}).
\newblock


\bibitem[Yao et~al\mbox{.}(2024)]%
        {yao2024tree}
\bibfield{author}{\bibinfo{person}{Shunyu Yao}, \bibinfo{person}{Dian Yu}, \bibinfo{person}{Jeffrey Zhao}, \bibinfo{person}{Izhak Shafran}, \bibinfo{person}{Tom Griffiths}, \bibinfo{person}{Yuan Cao}, {and} \bibinfo{person}{Karthik Narasimhan}.} \bibinfo{year}{2024}\natexlab{}.
\newblock \showarticletitle{Tree of thoughts: Deliberate problem solving with large language models}.
\newblock \bibinfo{journal}{\emph{Advances in Neural Information Processing Systems}}  \bibinfo{volume}{36} (\bibinfo{year}{2024}).
\newblock


\bibitem[Zhang et~al\mbox{.}(2024a)]%
        {zhang2024large}
\bibfield{author}{\bibinfo{person}{Chaoyun Zhang}, \bibinfo{person}{Shilin He}, \bibinfo{person}{Jiaxu Qian}, \bibinfo{person}{Bowen Li}, \bibinfo{person}{Liqun Li}, \bibinfo{person}{Si Qin}, \bibinfo{person}{Yu Kang}, \bibinfo{person}{Minghua Ma}, \bibinfo{person}{Qingwei Lin}, \bibinfo{person}{Saravan Rajmohan}, {et~al\mbox{.}}} \bibinfo{year}{2024}\natexlab{a}.
\newblock \showarticletitle{Large Language Model-Brained GUI Agents: A Survey}.
\newblock \bibinfo{journal}{\emph{arXiv preprint arXiv:2411.18279}} (\bibinfo{year}{2024}).
\newblock


\bibitem[Zhang et~al\mbox{.}(2023)]%
        {zhang2023appagent}
\bibfield{author}{\bibinfo{person}{Chi Zhang}, \bibinfo{person}{Zhao Yang}, \bibinfo{person}{Jiaxuan Liu}, \bibinfo{person}{Yucheng Han}, \bibinfo{person}{Xin Chen}, \bibinfo{person}{Zebiao Huang}, \bibinfo{person}{Bin Fu}, {and} \bibinfo{person}{Gang Yu}.} \bibinfo{year}{2023}\natexlab{}.
\newblock \showarticletitle{Appagent: Multimodal agents as smartphone users}.
\newblock \bibinfo{journal}{\emph{arXiv preprint arXiv:2312.13771}} (\bibinfo{year}{2023}).
\newblock


\bibitem[Zhang et~al\mbox{.}(2024c)]%
        {zhang2024aitz}
\bibfield{author}{\bibinfo{person}{Jiwen Zhang}, \bibinfo{person}{Jihao Wu}, \bibinfo{person}{Yihua Teng}, \bibinfo{person}{Minghui Liao}, \bibinfo{person}{Nuo Xu}, \bibinfo{person}{Xiao Xiao}, \bibinfo{person}{Zhongyu Wei}, {and} \bibinfo{person}{Duyu Tang}.} \bibinfo{year}{2024}\natexlab{c}.
\newblock \showarticletitle{Android in the zoo: Chain-of-action-thought for gui agents}.
\newblock \bibinfo{journal}{\emph{arXiv preprint arXiv:2403.02713}} (\bibinfo{year}{2024}).
\newblock


\bibitem[Zhang et~al\mbox{.}(2024d)]%
        {zhang2024mobileexperts}
\bibfield{author}{\bibinfo{person}{Jiayi Zhang}, \bibinfo{person}{Chuang Zhao}, \bibinfo{person}{Yihan Zhao}, \bibinfo{person}{Zhaoyang Yu}, \bibinfo{person}{Ming He}, {and} \bibinfo{person}{Jianping Fan}.} \bibinfo{year}{2024}\natexlab{d}.
\newblock \showarticletitle{MobileExperts: A Dynamic Tool-Enabled Agent Team in Mobile Devices}.
\newblock \bibinfo{journal}{\emph{arXiv preprint arXiv:2407.03913}} (\bibinfo{year}{2024}).
\newblock


\bibitem[Zhang et~al\mbox{.}(2024b)]%
        {zhang2024llamatouch}
\bibfield{author}{\bibinfo{person}{Li Zhang}, \bibinfo{person}{Shihe Wang}, \bibinfo{person}{Xianqing Jia}, \bibinfo{person}{Zhihan Zheng}, \bibinfo{person}{Yunhe Yan}, \bibinfo{person}{Longxi Gao}, \bibinfo{person}{Yuanchun Li}, {and} \bibinfo{person}{Mengwei Xu}.} \bibinfo{year}{2024}\natexlab{b}.
\newblock \showarticletitle{LlamaTouch: A Faithful and Scalable Testbed for Mobile UI Automation Task Evaluation}.
\newblock \bibinfo{journal}{\emph{arXiv preprint arXiv:2404.16054}} (\bibinfo{year}{2024}).
\newblock


\bibitem[Zhang and Zhang(2023)]%
        {zhang2023youautoui}
\bibfield{author}{\bibinfo{person}{Zhuosheng Zhang} {and} \bibinfo{person}{Aston Zhang}.} \bibinfo{year}{2023}\natexlab{}.
\newblock \showarticletitle{You only look at screens: Multimodal chain-of-action agents}.
\newblock \bibinfo{journal}{\emph{arXiv preprint arXiv:2309.11436}} (\bibinfo{year}{2023}).
\newblock


\end{thebibliography}
\bibliographystyle{ACM-Reference-Format}
\appendix\onecolumn
\newpage

\section{Additional LearnGUI Statistics}
\label{app:dataset_visualizations}

\begin{figure}[h]
	\centering
	\includegraphics[width=0.85\textwidth]{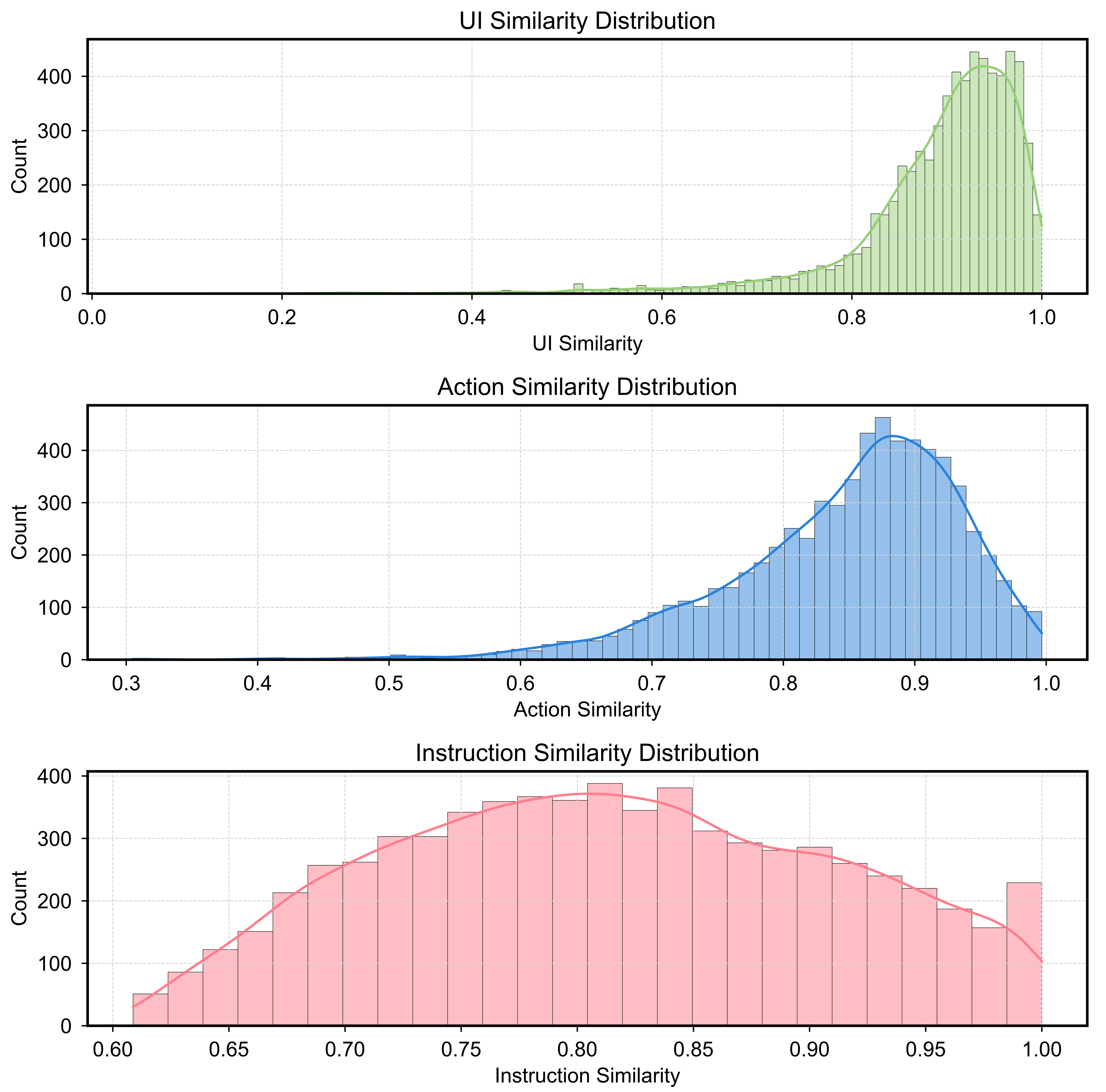}
	\caption{
		\textbf{Distribution of instruction, UI, and action similarity scores in LearnGUI-Offline.}
	\textmd{The histograms show the distribution of similarity scores across three dimensions: instruction similarity (top), UI similarity (middle), and action similarity (bottom). These distributions enable systematic analysis of how different types of similarity between demonstration and query tasks affect learning efficacy.}
	}
	\label{fig:similarity-distribution}
\end{figure}

Figure~\ref{fig:similarity-distribution} illustrates the distribution of similarity scores across different dimensions in the LearnGUI-Offline dataset, enabling systematic analysis of how different types of similarity between demonstration and query tasks affect learning efficacy.

\section{LearnAct Framework Details}
\label{app:method_details}

This section provides detailed descriptions of the components of our LearnAct framework, corresponding to the methods presented in Section 4 of the paper.

\subsection{DemoParser Prompts}
\label{app:demoparser_prompts}

We provide all of our prompt templates used in DemoParser for generating semantically descriptive action descriptions from demonstration data. These carefully designed prompts guide the vision-language model to produce structured knowledge that captures the essence of human demonstrations, as shown in Figures~\ref{fig:intermediate_prompt} and~\ref{fig:terminal_prompt}.

\begin{figure}[h]
\begin{response}[Intermediate Action Description]
\textbf{System Prompt:}

You are a mobile UI interaction analyst. Follow these rules:
1. Analyze the split-screen image (Before-action left, After-action right)
2. For click actions, a high-contrast red marker (white-bordered circle) shows the precise click location, with a green square surrounding it and a 'C' label at the top-right corner of the square indicating the click.
3. Output JSON with ONLY ONE 'action\_description' field in this exact format:
   "[On/In] [Screen Name], [Action Details], to [Purpose]"

\textbf{Action Types:}
- click [element] (e.g., 'Search button')
- swipe [up/down/left/right]
- type [text] in [field]
- press [back/home/enter]

\textbf{Validation Rules:}
1. Screen names should be 2-6 words
2. Keep purpose descriptions under 8 words
3. Never mention coordinates/IDs

\textbf{MEMORY RECORDING RULES:}
If the current screen contains information relevant to the user's instruction that needs to be remembered for future steps, include a Memory part in your action description. The format should be:
   "[On/In] [Screen Name], [Action Details], to [Purpose]. [Memory: important information for future steps]"

Memory should ONLY be added when:
1. The information is relevant to completing the user's instruction
2. The information will likely be needed in future steps
3. This specific information has NOT been recorded in previous action history entries

\textbf{Memory examples:}
1. For a travel planning task: On Travel Blog, click 'Bali Beach Guide', to read article. [Memory: Guide mentions Kuta Beach has surfing lessons for \$25/hour]
2. For a shopping task: In Product Details, click 'Add to Cart', to select item. [Memory: iPhone 13 Pro costs \$999 with 128GB storage]
3. For a note-taking task: On Weather App, swipe down forecast, to view weekend. [Memory: Saturday will be rainy with 80\% precipitation]

\textbf{Avoid using Memory for:}
1. Obvious UI changes that don't contain task-relevant information
2. Information already captured in previous action steps
3. Generic observations not specific to the user's task objective
\end{response}
\caption{
\textbf{Prompt template for intermediate action descriptions.}
\textmd{The template guides DemoParser to generate standardized descriptions for intermediate actions, including detailed rules for memory annotations that capture important information observed during task execution.}
}
\label{fig:intermediate_prompt}
\end{figure}

\begin{figure}[h]
\begin{response}[Terminal Action Description - Standard Completion]
\textbf{System Prompt for standard completion:}

Determine the final task status. Output rules:
1. Use ONLY ONE 'action\_description' field
2. Format: "[On/In] [Screen], complete task, [Reason]"

\textbf{Validation Rules:}
- Reason should be less than 10 words
- Screen name must match previous context

\textbf{Examples:}
1. Basic completion: On Payment Screen, complete task, successfully submit order
2. Failure case: In Search Results, cannot complete task, no nearby Vivo mobile phone stores found
\end{response}
\vspace{5mm}
\begin{response}[Terminal Action Description - With Answer]
\textbf{System Prompt for completion with answer:}

Determine the final task status with the given answer. Output rules:
1. Use ONLY ONE 'action\_description' field
2. Format: "[On/In] [Screen], complete task, the answer is [answer]"

\textbf{Validation Rules:}
- Screen name must match previous context
- Use the exact answer provided in the TASK\_COMPLETE action

\textbf{Examples:}
1. Answer is a price: On Checkout Screen, complete task, the answer is "\$299.9".
2. Answer is a list: On Payment Options Screen, complete task, the answer is "google pay, check out with affirm, add credit/debit card".
\end{response}
\caption{
\textbf{Prompt templates for terminal action descriptions.}
\textmd{The templates provide specific formats for both standard task completion and information retrieval tasks, ensuring consistent output structure across different task types.}
}
\label{fig:terminal_prompt}
\end{figure}

\subsection{ActExecutor Prompts}
\label{app:actexecutor_prompts}

We provide the prompt templates used by ActExecutor to make decisions based on current observations, action history, and demonstration knowledge. These prompts guide the vision-language model to select appropriate actions for task execution, as shown in Figure~\ref{fig:actexecutor_prompt}.

\begin{figure}[h]
\begin{response}[Task Execution Prompt]
\textbf{Role Definition:}

You are a smartphone assistant to help users complete tasks by interacting with apps. I will give you a screenshot of the current phone screen.

\textbf{Example Tasks:}
[Only when demonstrations are available]

Example 1: [Demonstration instruction]
Steps taken in this example:
Step-1: [Action] [Action Description]
Step-2: [Action] [Action Description]
...

\textbf{Background:}
This image is a phone screenshot. Its width is [width] pixels and its height is [height] pixels.
The user's instruction is: [instruction]

\textbf{History operations:}
[Only when action history is available]

Before reaching this page, some operations have been completed. You need to refer to the completed operations to decide the next operation. These operations are as follow:
Step-1: [Action] [Action Description]
Step-2: [Action] [Action Description]
...

\textbf{Response requirements:}
Now you need to combine all of the above to decide just one action on the current page. You must choose one of the actions below:

\texttt{"SWIPE[UP]"}: Swipe the screen up.
\texttt{"SWIPE[DOWN]"}: Swipe the screen down.
\texttt{"SWIPE[LEFT]"}: Swipe the screen left.
\texttt{"SWIPE[RIGHT]"}: Swipe the screen right.
\texttt{"CLICK[x,y]"}: Click the screen at the coordinates (x, y). x is the pixel from left to right and y is the pixel from top to bottom
\texttt{"TYPE[text]"}: Type the given text in the current input field.
\texttt{"PRESS\_BACK"}: Press the back button.
\texttt{"PRESS\_HOME"}: Press the home button.
\texttt{"PRESS\_ENTER"}: Press the enter button.
\texttt{"TASK\_COMPLETE[answer]"}: Mark the task as complete. If the instruction requires answering a question, provide the answer inside the brackets. If no answer is needed, use empty brackets \texttt{"TASK\_COMPLETE[]"}.

\textbf{Response Example:}
Your output should be a string and nothing else, containing only the action type you choose from the list above.
For example:
\texttt{"SWIPE[UP]"}
\texttt{"CLICK[156,2067]"}
\texttt{"TYPE[Rome]"}
\texttt{"PRESS\_BACK"}
\texttt{"PRESS\_HOME"}
\texttt{"PRESS\_ENTER"}
\texttt{"TASK\_COMPLETE[1h30m]"}
\texttt{"TASK\_COMPLETE[]"}
\end{response}
\caption{
\textbf{Task execution prompt template.}
\textmd{This comprehensive prompt directs ActExecutor to generate actions based on current observations, action history, and retrieved demonstrations, with explicit formatting requirements to ensure consistent action outputs.}
}
\label{fig:actexecutor_prompt}
\end{figure}

\subsection{Algorithm Details}
\label{app:algorithm_details}

We provide the detailed algorithms for the DemoParser and ActExecutor components of our LearnAct framework, which are the core computational processes enabling knowledge extraction and task execution.

\begin{algorithm}
\caption{DemoParser Knowledge Generation Process}
\label{alg:demoparser_appendix}
\begin{algorithmic}[1]
\Require Demonstration dataset $D = \{(i_k, s_k, a_k)_{k=1}^N\}$ where $i_k$ is instruction, $s_k$ is screenshot sequence, $a_k$ is action sequence
\Ensure Knowledge base $K$ with semantically descriptive action descriptions
\State $K \leftarrow \emptyset$ \Comment{Initialize empty knowledge base}
\For{each demonstration $(i, s, a)$ in $D$}
    \State $d \leftarrow \emptyset$ \Comment{Initialize empty description sequence}
    \For{$j = 1$ to $|a|$}
        \If{$j < |a|$} \Comment{Intermediate action}
            \State Create visualization of action $a_j$ with before-after screenshots from $s_j$ and $s_{j+1}$
            \State $h \leftarrow $ Previous action descriptions $\{d_1, d_2, \ldots, d_{j-1}\}$
            \State $d_j \leftarrow $ GenerateDescription$(i, a_j, visualization, h)$ using prompt format detailed in Appendix~\ref{app:demoparser_prompts}
            \State $d_j$ follows format: "[On/In] [Screen], [Action], to [Purpose]" with optional memory
        \Else \Comment{Terminal action}
            \State $h \leftarrow $ Complete action history $\{d_1, d_2, \ldots, d_{|a|-1}\}$
            \State $d_{|a|} \leftarrow $ GenerateFinalDescription$(i, s_{|a|}, h, a_{|a|})$ using prompt detailed in Appendix~\ref{app:demoparser_prompts}
            \State $d_{|a|}$ follows format: "[On/In] [Screen], complete task, [Reason/Answer]"
        \EndIf
        \State Add $d_j$ to description sequence $d$
    \EndFor
    \State Add $(i, a, d)$ to knowledge base $K$
\EndFor
\State \Return $K$
\end{algorithmic}
\end{algorithm}

\begin{algorithm}
\caption{ActExecutor Task Execution Process}
\label{alg:actexecutor_appendix}
\begin{algorithmic}[1]
\Require User instruction $i$, Knowledge base $K$, Maximum steps $T$
\Ensure Task execution trajectory
\State $t \leftarrow 0$ \Comment{Initialize time step}
\State $h \leftarrow \emptyset$ \Comment{Initialize action history}
\State $\mathcal{D} \leftarrow$ KnowSeeker$(i, K)$ \Comment{Retrieve relevant demonstrations}
\While{$t < T$ and not IsTaskComplete}
    \State $o_t \leftarrow$ GetObservation() \Comment{Obtain current screenshot}
    \State $P_t \leftarrow$ ConstructPrompt$(i, o_t, h, \mathcal{D})$ \Comment{Construct decision prompt}
    \State $a_t \leftarrow f_{LLM}(P_t)$ \Comment{Generate action via LLM}
    \State $d_t \leftarrow$ GenerateDescription$(i, a_t, o_t, h)$ \Comment{Generate action description}
    \State $h \leftarrow h \cup \{(a_t, d_t)\}$ \Comment{Update action history}
    \State ExecuteAction$(a_t)$ \Comment{Execute action in environment}
    \State $t \leftarrow t + 1$ \Comment{Increment time step}
\EndWhile
\State \Return $\{(a_0, d_0), (a_1, d_1), \ldots, (a_{t-1}, d_{t-1})\}$
\end{algorithmic}
\end{algorithm}

\section{Additional Experimental Results and Analyses}
\label{app:experimental_results}

This section provides additional experimental results and analyses that supplement the findings presented in Section 5 of the paper.

\subsection{Online Performance Comparisons}
\label{app:online_visualizations}

\begin{figure}[h]
	\centering
	\includegraphics[width=0.9\textwidth]{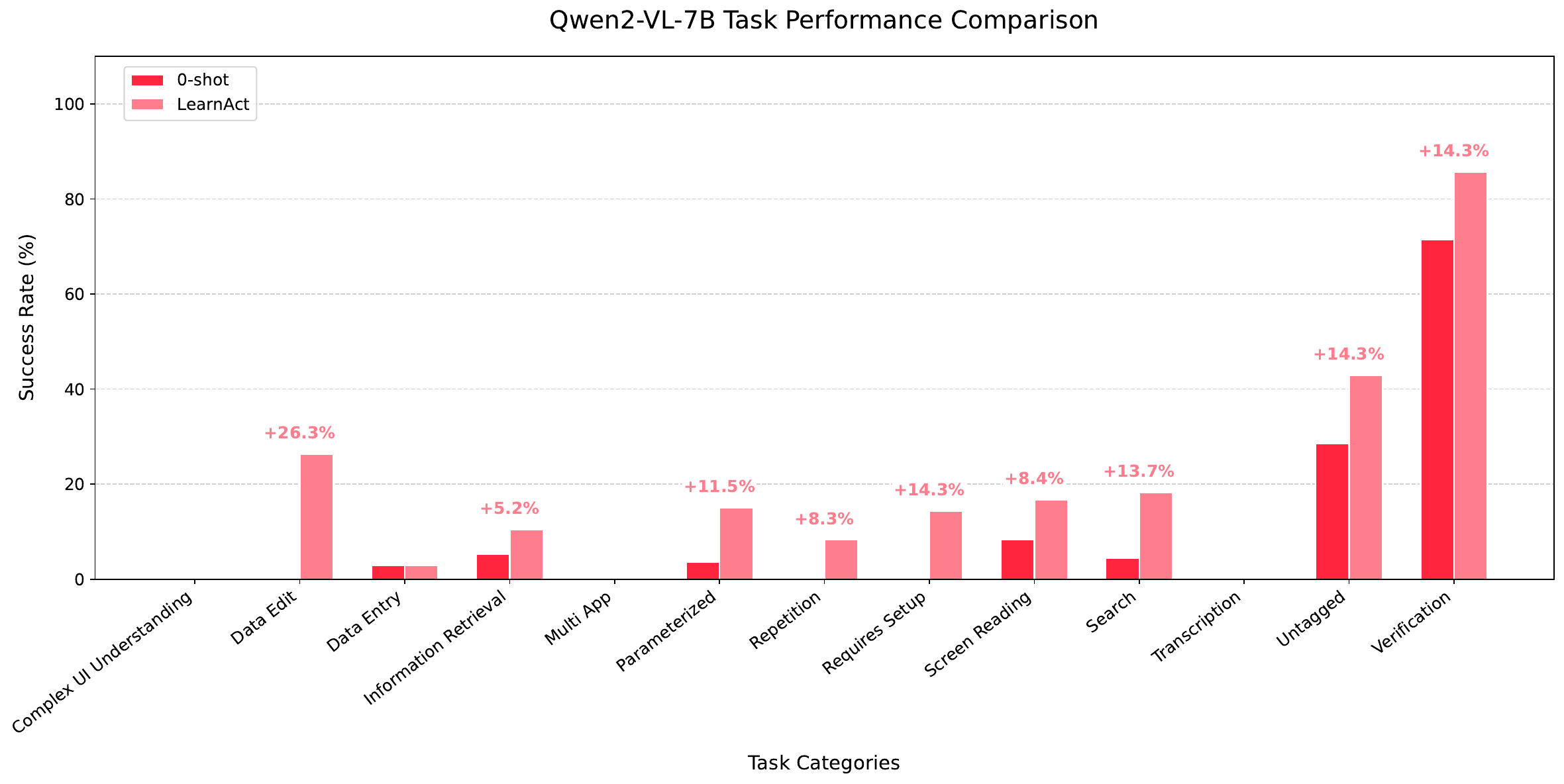}
	\caption{
		\textbf{Detailed performance comparison of Qwen2-VL-7B with and without LearnAct on LearnGUI-Online.}
	\textmd{The figure shows the task success rates of Qwen2-VL-7B baseline versus Qwen2-VL-7B enhanced with LearnAct across different task dimensions in the LearnGUI-Online benchmark.}
	}
	\label{fig:qwen-learnact-online-comparison}
\end{figure}

\begin{figure}[h]
	\centering
	\includegraphics[width=0.9\textwidth]{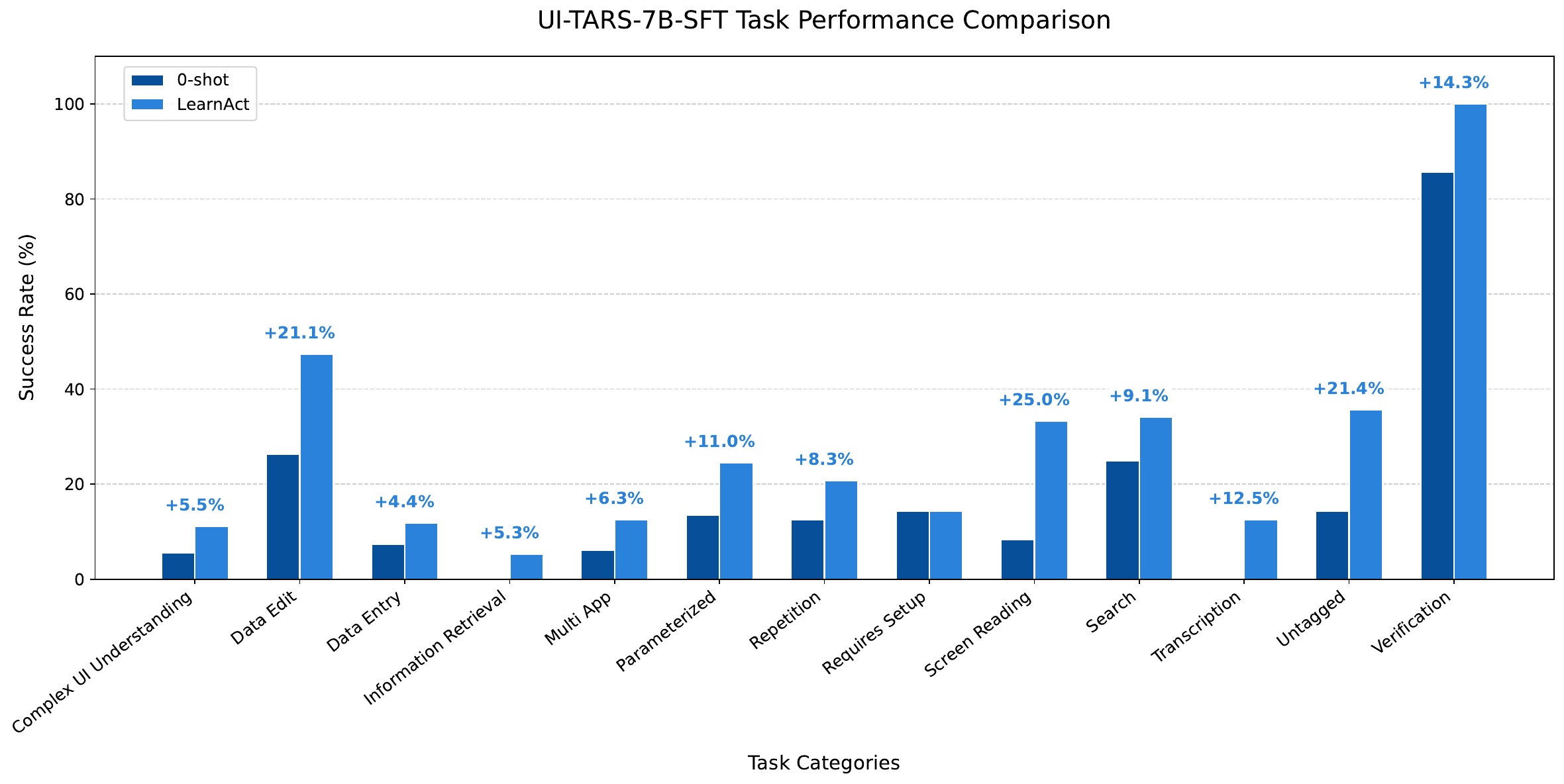}
	\caption{
		\textbf{Detailed performance comparison of UI-TARS-7B-SFT with and without LearnAct on LearnGUI-Online.}
	\textmd{The figure presents a comprehensive breakdown of task success rates for UI-TARS-7B-SFT baseline versus UI-TARS-7B-SFT enhanced with LearnAct across multiple task dimensions in the LearnGUI-Online benchmark. }
	}
	\label{fig:uitars-learnact-online-comparison}
\end{figure}

Figures~\ref{fig:qwen-learnact-online-comparison} and~\ref{fig:uitars-learnact-online-comparison} provide detailed comparisons of model performance with and without LearnAct enhancement in online evaluation scenarios.

\subsection{Case Studies of LearnAct Online Experiments}
\label{app:case_studies}

We present three detailed case studies from our online experiments to provide concrete examples of how LearnAct leverages demonstration knowledge to solve tasks in unseen mobile applications. These case studies highlight different scenarios where demonstration knowledge proves particularly beneficial for task execution.
\begin{figure}[ht]
\centering
\includegraphics[width=0.8\textwidth]{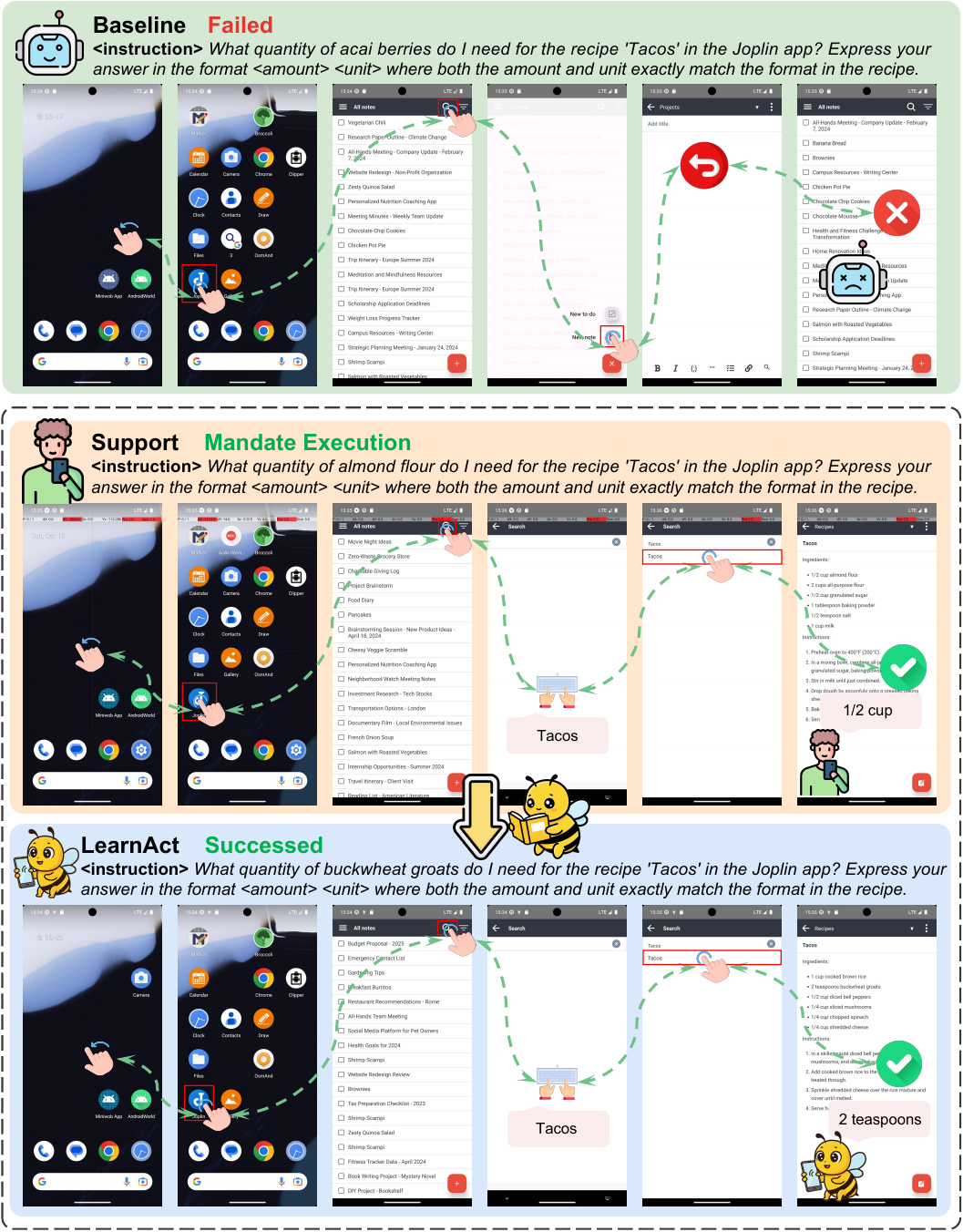}
\caption{
\textbf{UI-TARS-7B-SFT with LearnAct vs. Baseline in NotesRecipeIngredientCount Task.}
\textmd{Task template: "What quantity of \{ingredient\} do I need for the recipe '\{title\}' in the Joplin app? Express your answer in the format <amount> <unit> without using abbreviations."}
}
\label{fig:case-study1}
\end{figure}

\begin{figure}[ht]
\centering
\includegraphics[width=0.7\textwidth]{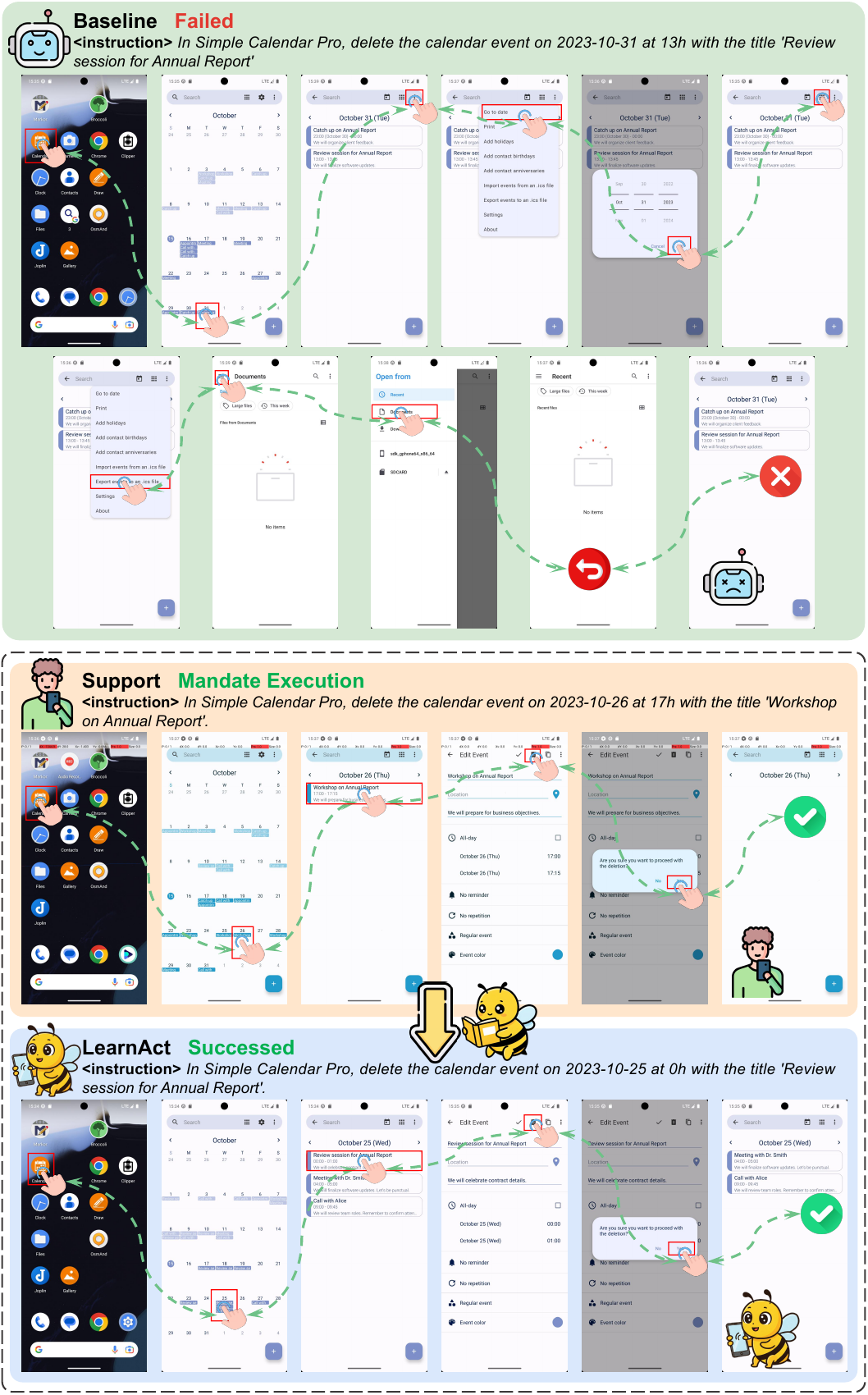}
\caption{
\textbf{UI-TARS-7B-SFT with LearnAct vs. Baseline in SimpleCalendarDeleteOneEvent Task.}
\textmd{Task template: "In Simple Calendar Pro, delete the calendar event on \{year\}-\{month\}-\{day\} at \{hour\}h with the title '\{event\_title\}'"}
}
\label{fig:case-study2}
\end{figure}

\begin{figure}[ht]
\centering
\includegraphics[width=0.7\textwidth]{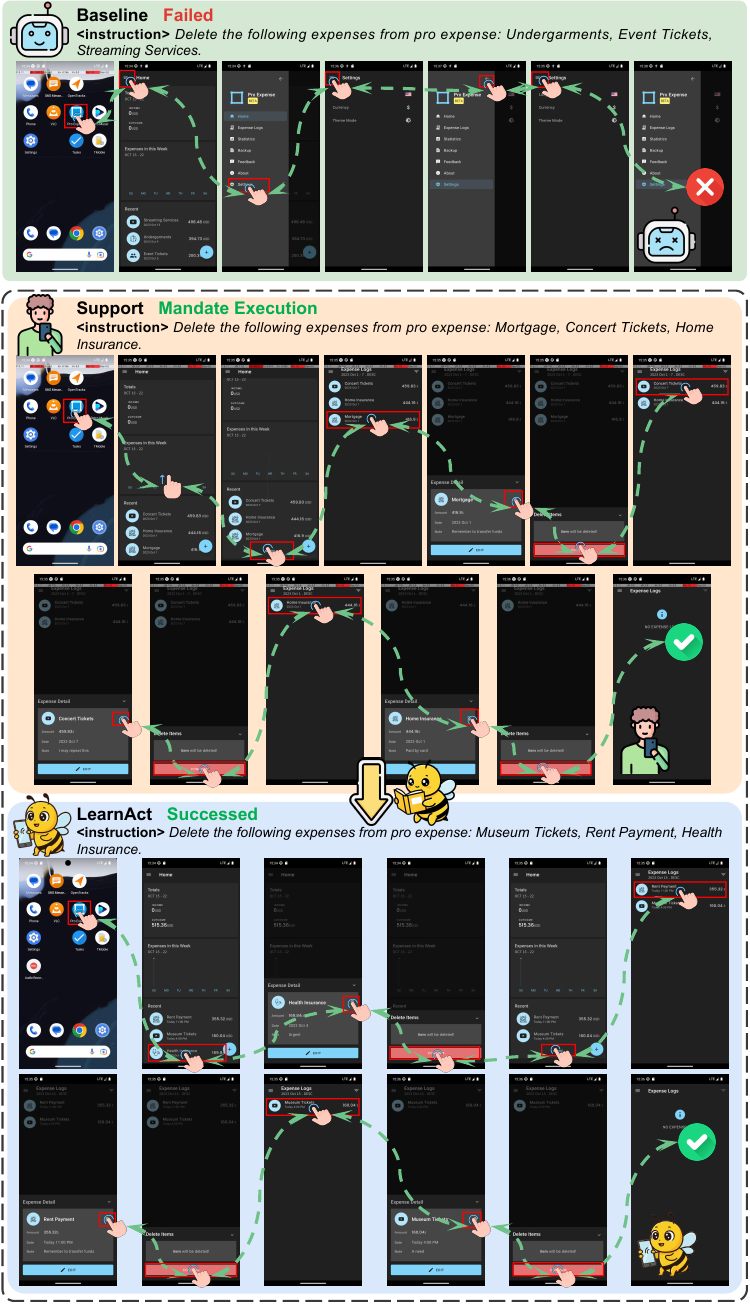}
\caption{
\textbf{Qwen2-VL-7B with LearnAct vs. Baseline in ExpenseDeleteMultiple Task.}
\textmd{Task template: "Delete the following expenses from arduia pro expense: \{expenses\}."}
}
\label{fig:case-study3}
\end{figure}

\end{document}